\pdfoutput=1
\documentclass[aps,prd,amsmath,floats,floatfix, twocolumn,
superscriptaddress,nofootinbib,showpacs,longbibliography]{revtex4-1}
\usepackage{aas_macros}
\usepackage[T1]{fontenc}
\usepackage[utf8]{inputenc}
\usepackage{txfonts}
\usepackage{verbatim}
\usepackage[dvipsnames, usenames]{xcolor}
\definecolor{linkcolor}{rgb}{0.0,0.3,0.5}
\usepackage[hypertexnames=false, unicode, colorlinks=true, linkcolor=linkcolor,
citecolor=linkcolor, filecolor=linkcolor,urlcolor=linkcolor,
pdfusetitle]{hyperref}
\usepackage[all]{hypcap}
\usepackage{graphicx}
\usepackage{xspace}
\usepackage{amssymb}
\usepackage{microtype}
\usepackage{subfigure}

\usepackage{url}

\usepackage[english]{babel}
\usepackage{blindtext}

\usepackage[normalem]{ulem} 
\usepackage{bm} 
\usepackage{mathrsfs}
\usepackage{xspace} 
\usepackage{fontawesome}

\DeclareMathAlphabet{\mathpzc}{OT1}{pzc}{m}{it}

\usepackage[nomessages]{fp}

\newcommand{\sk}[1]{}
\defcitealias{Islam:2025drw}{IW25}

\newcommand{\rapster}{\textcolor{linkcolor}{\texttt{rapster}}}
\newcommand{\gwGenealogy}{\textcolor{linkcolor}{\texttt{gwGenealogy}}}
\newcommand{\gwModel}{\textcolor{linkcolor}{\texttt{gwModel\_flow\_prec}}}
\newcommand{\HLZ}{\textcolor{linkcolor}{\texttt{HLZ}}}

\begin{document}

\title{Kick matters: The impact of a new recoil model on the retention of hierarchical black-hole remnants in globular clusters}

\author{Tousif Islam}
\email{tousifislam@ucsb.edu}
\affiliation{Kavli Institute for Theoretical Physics,University of California Santa Barbara, Kohn Hall, Lagoon Rd, Santa Barbara, CA 93106}

\author{Digvijay Wadekar}
\affiliation{\mbox{Weinberg Institute, University of Texas at Austin, Austin, TX 78712, USA}}
\affiliation{Department of Physics and Astronomy, Johns Hopkins University, 3400 N. Charles Street, Baltimore, Maryland, 21218, USA}

\author{Konstantinos Kritos}
\affiliation{Department of Physics and Astronomy, Johns Hopkins University, 3400 N. Charles Street, Baltimore, Maryland, 21218, USA}

\hypersetup{pdfauthor={Islam et al.}}

\date{\today}

\begin{abstract}
In globular clusters, hierarchical mergers are among the most promising pathways to forming massive black holes such as GW231123. A key factor determining whether a merger-remnant black hole will be retained in these environments and thus participate in subsequent hierarchical mergers is the recoil kick velocity. Analytic models for the recoil velocity are currently employed in nearly all population-synthesis frameworks.
We instead use a state-of-the-art recoil-kick model \gwModel{}\footnote{\href{https://github.com/tousifislam/gwModels}{https://github.com/tousifislam/gwModels}\label{gwModel}}~\cite{Islam:2025drw} developed from a combination of numerical-relativity and black-hole–perturbation-theory data, together with data-driven techniques such as normalizing flows and the post-Newtonian structure of the kick.  
Employing both back-of-the-envelope estimates and detailed N-body as well as semi-analytical cluster simulations, we show that \gwModel{} leads to a noticeable increase in the retention probability of hierarchical-merger remnants compared to the previously used analytic model and changes the mass and spin distribution of the black holes formed through hierarchical mergers. Additionally, we discuss the implications of our results in the context of massive binaries such as GW231123.
\end{abstract}

\maketitle

\section{Introduction}
\label{sec:intro}
One of the consequences of binary black-hole (BBH) mergers in general relativity (GR) is the recoil kick imparted to the remnant black hole (BH)~\cite{Baker:2006vn,Campanelli:2007cga,Nakano:2010kv}. This kick arises from the rapid change in linear momentum during the merger and can reach velocities as large as ${\sim}5000\,$km\,s$^{-1}$~\cite{Campanelli:2007cga,Bruegmann:2007bri, Campanelli:2007ew, Choi:2007eu, Dain:2008ck,Gonzalez:2006md,Gonzalez:2007hi,Healy:2008js,Healy:2022jbh}. Over the past two decades, this phenomenon has been studied in detail using a combination of numerical relativity (NR)~\cite{Baker:2006vn, Baker:2007gi, Baker:2008md,Herrmann:2006cd,Lousto:2007db,Herrmann:2007ac,Herrmann:2007ex,Herrmann:2007cwl,Holley-Bockelmann:2007hmm,Jaramillo:2011re,Koppitz:2007ev,Lousto:2008dn,Lousto:2010xk,Schnittman:2007ij,Sopuerta:2006et,Pollney:2007ss,Rezzolla:2010df,Lousto:2011kp,Lousto:2012gt,Lousto:2012su,Miller:2008en,Tichy:2007hk,Zlochower:2010sn,Healy:2014yta,Lousto:2009mf}, black-hole perturbation theory (BHPT)~\cite{Nakano:2010kv,Sundararajan:2010sr,Islam:2023mob,Hughes:2004ck,Price:2013paa,Price:2011fm}, and post-Newtonian (PN) approximations~\cite{Blanchet:2005rj,Sopuerta:2006wj,Sopuerta:2006et,Favata:2004wz,Fitchett:1983qzq,Fitchett:1984qn,Wiseman:1992dv,Kidder:1995zr}, leading to a rich phenomenology. Building on these developments, several fast approximate models for the recoil velocity have emerged~\cite{Zlochower:2015wga,Baker:2008md,Lousto:2008dn,Lousto:2010xk,Lousto:2012gt,Lousto:2012su,vanMeter:2010md,Healy:2014yta,Sundararajan:2010sr,Islam:2023mob,Varma:2019csw,Varma:2018aht,Merritt:2004xa,Kidder:1995zr,Sperhake:2019wwo,Islam:2025drw}, and these are now routinely employed in gravitational-wave (GW) data analysis~\cite{Mahapatra:2021hme,Varma:2022pld,Islam:2023zzj,CalderonBustillo:2022ldv}.

One of the key astrophysical consequences of BBH mergers is the possibility of \emph{hierarchical mergers}~\cite{Gerosa:2016vip,Borchers:2025sid}, in which black holes (BHs) produced as merger remnants subsequently merge again. The initial, or first-generation (1G), BH population is expected to originate from stellar collapse and typically has masses $\lesssim 50\,M_{\odot}$~\cite{Spera:2017fyx,Franciolini:2024vis}. However, when two 1G BHs merge, the resulting remnant, a second-generation (2G) BH, can exceed this mass scale. Subsequent mergers between 2G and 1G BHs, or between two 2G BHs, can then produce third-generation (3G) BHs. This process can of course continue further. This dynamical assembly mechanism is widely considered a promising pathway for forming intermediate-mass black holes (IMBHs) seeds and, ultimately, supermassive black holes (SMBHs)~\cite{Holley-Bockelmann:2007hmm,Berti:2012zp,Gultekin:2004pm,Gerosa:2016vip,Borchers:2025sid,Gerosa:2021hsc,Gerosa:2017kvu,Gerosa:2021hsc,Baibhav:2020xdf,Bouffanais:2019nrw}. Recently, several GW events, such as GW190521 and GW231123, have been interpreted as involving BHs with masses of order $100\,M_{\odot}$ or higher. These observations have renewed interest in hierarchical-merger scenarios.

However, it is not guaranteed that all remnant BHs will participate in hierarchical mergers within clusters or galactic nuclei. A key ingredient in determining whether hierarchical mergers can occur is the interplay between the recoil kick velocity imparted to the remnant BH at merger and the escape velocity of the host environment~\cite{Holley-Bockelmann:2007hmm,Berti:2012zp,Gultekin:2004pm,Gerosa:2016vip,Borchers:2025sid,Gerosa:2021hsc,Gerosa:2017kvu,Gerosa:2021hsc,Baibhav:2020xdf,Bouffanais:2019nrw}. If the kick velocity is significantly larger than the escape velocity, the remnant will be ejected from the host system (e.g., a globular cluster). Conversely, if the kick velocity is smaller than the escape velocity, the remnant BH will remain bound, albeit potentially displaced from the merger site. It can then return in the center through efficient dynamical friction on a timescale that is smaller than the stellar relaxation time by a factor $\sim m_\star/m_{\rm BH}$, and the BH can subsequently participate in additional mergers. 

In the past, hierarchical mergers have been studied extensively using back-of-the-envelope calculations applied to simplified cluster models, often incorporating varying assumptions about cluster astrophysics~\cite{Holley-Bockelmann:2007hmm,Berti:2012zp,Gultekin:2004pm,Gerosa:2016vip,Borchers:2025sid,Gerosa:2021hsc,Gerosa:2017kvu,Gerosa:2021hsc,Baibhav:2020xdf,Bouffanais:2019nrw}. In parallel, significant effort has gone into modeling cluster dynamics, including hierarchical mergers~\cite{Gerosa:2016vip,Borchers:2025sid,Zevin:2022bfa,Torniamenti:2024uxl}, using $N$-body simulations~\cite{ArcaSedda:2023mlv}, Monte Carlo methods~\cite{Rodriguez:2019huv}, and semi-analytical frameworks~\cite{Kritos:2022non}. Various codebases such as \texttt{Cluster Monte Carlo}~\cite{Rodriguez:2019huv,Kremer:2019iul}, McFacts~\cite{McKernan:2024kpr}, rapster \cite{Kritos:2022non}, fastcluster~\cite{Mapelli:2021Gyv}, \texttt{BBHDynamics}~\cite{Antonini:2019ulv}, \texttt{BPOP}~\cite{Sedda:2021vjh} have been developed to model the demographics of compact objects originating from dynamical channels. All these studies have almost invariably relied on analytic recoil-kick models \cite{Lousto:2008dn,Lousto:2010xk,Lousto:2012gt,Lousto:2012su,Gonzalez:2007hi}, which were developed years ago from a relatively small NR dataset that was heavily biased toward equal-mass, aligned-spin binaries.

~\citet{Islam:2025drw} (hereafter \citetalias{Islam:2025drw})  recently presented an updated model \gwModel{}$^{\ref{gwModel}}$ for the recoil kick based on a significantly larger dataset constructed from a combination of NR simulations and BHPT. Their model incorporates analytical insights from PN theory and leverages data-driven techniques such as normalizing flows to produce more accurate recoil-velocity model than the analytic \HLZ{} model. Furthermore, \citetalias{Islam:2025drw} demonstrated that previously used analytic models tend to overestimate the kick in many regions of parameter space, which would artificially suppress the predicted retention efficiency of BHs undergoing hierarchical mergers. Previously, the predicted large recoil kicks disfavored hierarchical mergers as an efficient pathway for growing massive black holes in dense stellar clusters, motivating the community to instead explore alternative runaway formation channels — including successive tidal disruption events (e.g., Ref.~\cite{Rizzuto:2022fdp,Kritos:2025bby}) and repeated stellar collisions (e.g., Ref.~\cite{Paiella:2025tia}) — as viable mechanisms for producing massive black holes in this environment. The goal of our paper is to use our updated recoil kick model to re-examine hierarchical-merger pathways and quantify the resulting astrophysical implications.

To illustrate how the updated recoil-kick model can significantly affect several astrophysical conclusions from previous studies, we employ a range of dynamical population synthesis methods. First, we conduct a series of semi-analytical simulations that vary initial mass and spin distributions, as well as the binary pairing functions, to explore how many black holes can form and how massive they can become through hierarchical mergers under different recoil-kick prescriptions. Next, we perform approximately 7500 detailed astrophysical cluster simulations using the semi-analytical cluster evolution code \texttt{rapster}. By adopting a representative set of cluster properties with different random seeds, we highlight how the landscape of hierarchical mergers is altered when using the new recoil model compared to older prescriptions.

The remainder of this paper is organized as follows. In Sec.~\ref{sec:model}, we outline our approach to modeling the recoil kick. Section~\ref{sec:simple_calculations} presents a series of simplified numerical experiments using our baseline model, aimed at elucidating the effects of model uncertainties, the initial black hole population, and binary pairing functions on the resulting masses and spins of black holes formed through hierarchical mergers. 
These studies make use of the \gwGenealogy{} package,\footnote{\href{https://github.com/tousifislam/gwGenealogy}{https://github.com/tousifislam/gwGenealogy}} which we have developed as part of this work to offer a straightforward, unified, and publicly accessible framework for studying hierarchical mergers.

\begin{figure}
    \centering
    \includegraphics[width=\columnwidth]{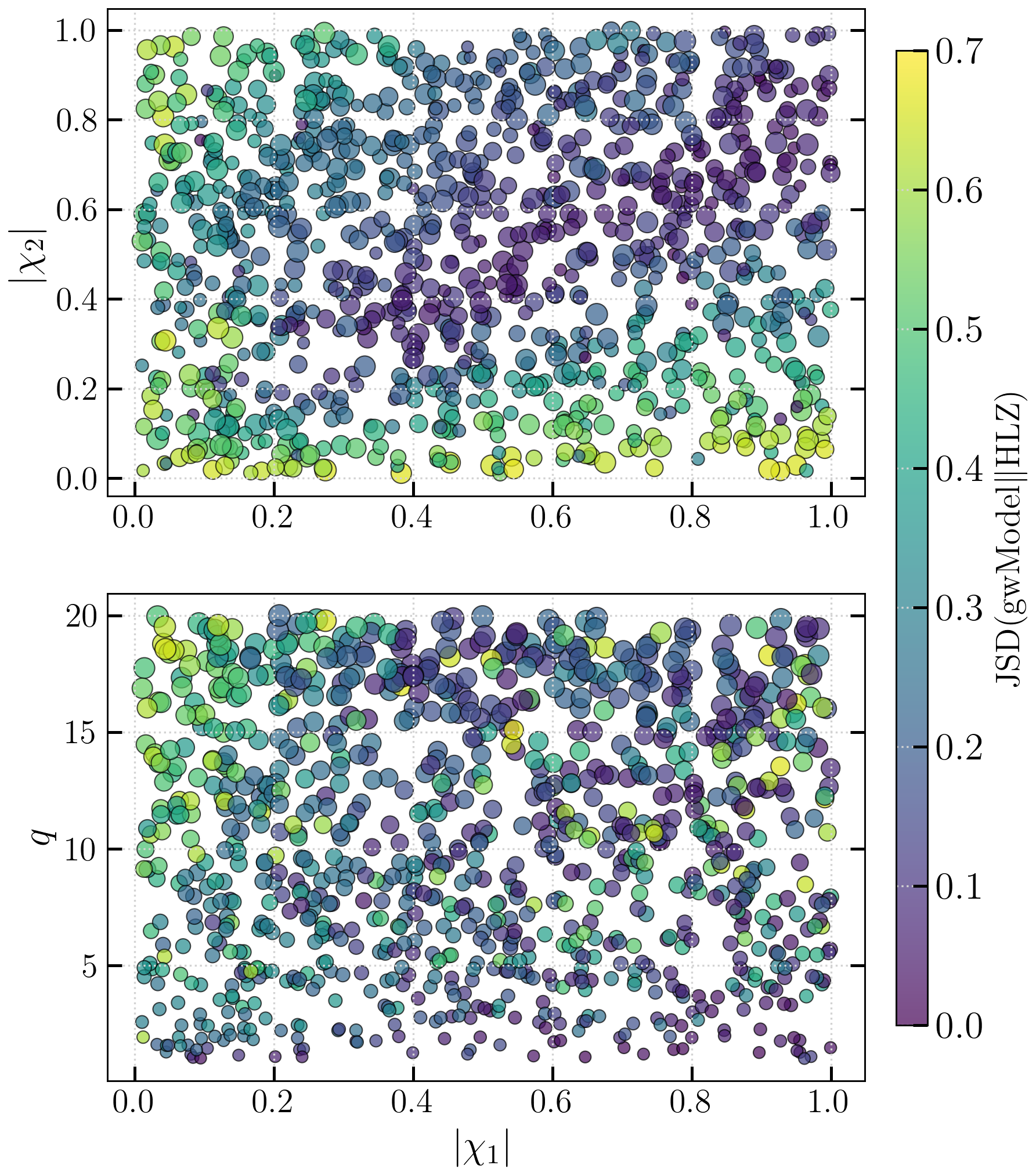}
    \caption{We show the Jensen--Shannon divergence (JSD; color-coded) between the kick-velocity distributions predicted by the new \gwModel{} model and the widely-used analytic model \HLZ{}. The kick distributions for each point in the $q, |\chi_1|, |\chi_2|$ space is generated by drawing 5000 spin orientation angles from an isotropic distribution. The symbol size scales with the mass ratio: smaller circles correspond to smaller $q$, while larger circles correspond to larger $q$. There is substantial discrepancies between the two models when one or both spin magnitudes are small and/or when the mass ratio becomes increasingly asymmetric. More details are in Section~\ref{sec:model}.}
    \label{fig:jsd_gwModel_vs_HLZ}
\end{figure}

Building upon the insights from these simplified calculations, we next move to more detailed and realistic cluster simulations in Sec.~\ref{sec:rapster}, exploring various recoil-kick prescriptions. These simulations use the \rapster{} package\footnote{\url{https://github.com/Kkritos/Rapster}}~\cite{Kritos:2022ggc} and \textcolor{linkcolor}{\texttt{CMC}}~\cite{Kremer:2019iul}. Throughout Secs.~\ref{sec:simple_calculations} and \ref{sec:rapster}, we relate our findings to the population of massive black holes associated with the GW231123 event. In Section~\ref{sec:discussion}, we discuss the broader implications of our results and outline potential directions for future work.

\section{Model for the recoil velocity}
\label{sec:model}
A BBH merger is characterized by two black holes with masses $m_{1,2}$ (with $m_1 \geq m_2$), dimensionless spin magnitudes $\chi_{1,2}\in [0,1]$, and four spin-orientation angles $\{\theta_1, \theta_2, \phi_1, \phi_2\}$. The angles $\{\theta_1, \theta_2\}\in [0,\pi]$ denote the tilt of each spin vector relative to the orbital angular-momentum axis, while $\{\phi_1, \phi_2\}\in [0,2\pi]$ specify the in-plane spin angles projected onto the orbital plane. The remnant black hole has mass $m_f$ and dimensionless spin $\chi_f$, and the magnitude of the recoil (kick) velocity is denoted by $v_f$. In this work, we focus exclusively on the magnitude of the kick, as the kick direction is not especially relevant for hierarchical mergers in clusters which has isotropic symmetry.

\begin{figure*}
    \centering
    \includegraphics[width=0.95\textwidth]{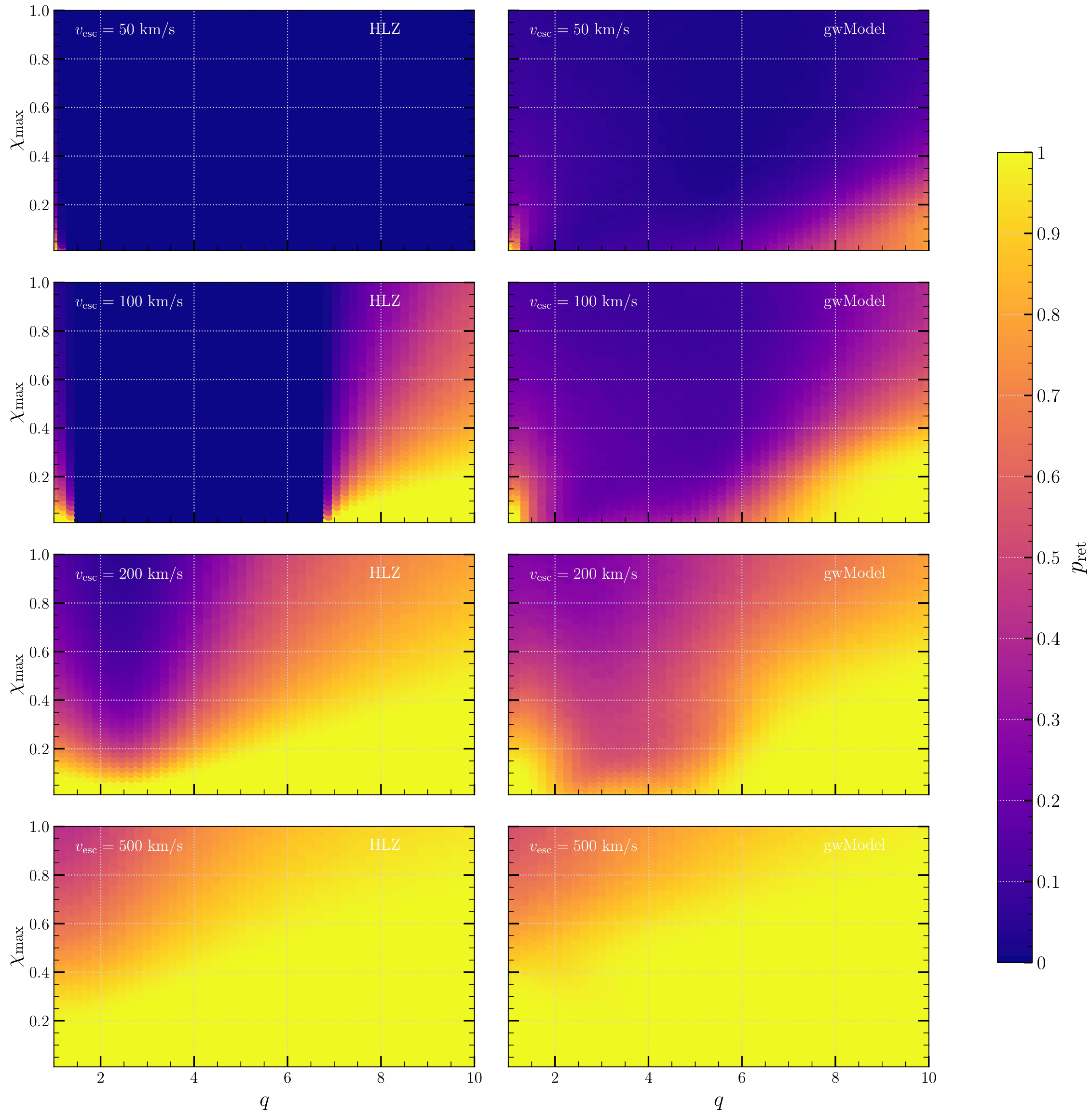}
    \caption{Probability of retaining the merger remnant of a BBH (shown by the colorbar) as a function of the BBH mass ratio and the maximum black-hole spins, shown for different cluster escape speeds: 50$-$500 km s$^{-1}$ (increasing from top to bottom). The left columns show the results obtained using the \HLZ{} recoil-kick model, while the right columns show the results for the \gwModel{}. The black-hole spins are drawn from a uniform distribution in the range $[0, \chi_{\max}]$ and spin angles are chosen isotropically. Details are in Section~\ref{sec:pret}.}
    \label{fig:pret_q_chimax_grid}
\end{figure*}

To compute recoil kick velocities, essentially all $N$-body cluster-simulation codes and semi-analytical frameworks rely on the semi-analytical fitting formulae developed in Refs.~\cite{Lousto:2008dn,Lousto:2010xk,Lousto:2012gt,Lousto:2012su,Gonzalez:2007hi}, which we collectively refer to as the \HLZ{} model (after the authors).
More recently, some theoretical studies~\cite{GalvezGhersi:2020fvh,Borchers:2025sid} have employed NR surrogate models, which offer improved accuracy relative to \HLZ{}. However, these surrogate models are restricted to mass ratios $1 \leq q \leq 4$ for precessing systems and $1 \leq q \leq 8$ for aligned-spin systems, and they are approximately three orders of magnitude slower than the semi-analytical \HLZ{} fits, limiting their applicability in large-scale cluster simulations. 
To overcome these challenges, \citetalias{Islam:2025drw} have recently developed \gwModel{}, a normalizing-flow–based model that accurately captures the distribution of recoil kick velocities arising from varying spin orientations. 

To explicitly quantify the differences between the new recoil model \gwModel{} and the earlier prescription \HLZ{}, we compute recoil kick velocity distributions for 1000 randomly selected combinations of the mass ratio $q \in [1,20]$ and spin magnitudes $|\chi_{1,2}| \in [0,1]$. For each combination of mass ratio and spin magnitudes, we draw 5000 spin orientation angles from an isotropic distribution i.e. $\cos\theta_{1,2} \in \mathcal{U}(-1)$ and $\phi_{1,2}\in \mathcal{U}(0,2\pi)$. 

To measure the discrepancy between the kick velocity distributions predicted by the two models, we compute the Jensen--Shannon divergence (JSD). 
The JSD between two probability distributions $p$ and $q$ is defined as~\cite{Js61115}
\begin{equation}
\mathrm{JSD}(p_1||p_2) = \tfrac{1}{2}\mathrm{KL}(p_1||m) + \tfrac{1}{2}\mathrm{KL}(p_2||m),
\end{equation}
where $m=\tfrac{1}{2}(p+q)$ is their pointwise mean and $\mathrm{KL}$ denotes the Kullback--Leibler (KL) divergence~\cite{Js61115} 
\begin{equation}
\mathrm{KL}(X||Y)
=
\int X(x)\,\log\!\left(\frac{X(x)}{Y(x)}\right)\,dx.
\end{equation}
The JSD is symmetric and bounded, $0 \le \mathrm{JSD}(p||q) \le \log 2 \approx 0.693$. Values $\mathrm{JSD} \lesssim 0.02$ indicate virtually indistinguishable distributions, while $\mathrm{JSD} \gtrsim 0.1$ signifies substantial differences; intermediate values correspond to moderately distinct distributions.

In Figure~\ref{fig:jsd_gwModel_vs_HLZ}, we show the JSD as a function of the binary parameter space. We find that the kick velocity distributions predicted by \gwModel{} and \HLZ{} differ significantly across large regions of parameter space. In particular, when one or both spin magnitudes are small, the JSD increases markedly, indicating substantial discrepancies between the two models. In contrast, the two prescriptions yield more similar kick distributions for nearly equal-mass binaries except for slowly spinning binaries. 
As the mass ratio becomes increasingly asymmetric, the differences between the models grow. Note that separate versions of Fig.~\ref{fig:jsd_gwModel_vs_HLZ} for the low and high mass ratio cases were presented in \citetalias{Islam:2025drw} (see their Figs.~7 and~8). They also show that \gwModel{} agrees much better with NR surrogate models compared to \HLZ{} for the low mass ratio case (which is the regime where the NR surrogate model is expected to be more accurate compared to \HLZ{}).

\section{Back-of-the-envelope calculations}
\label{sec:simple_calculations}
We use \gwGenealogy{} to perform a series of experiments to build intuition about hierarchical mergers under simplified astrophysical assumptions. Each experiment is carried out twice: once using the \HLZ{} recoil model and once using the \gwModel{} recoil model.

For the initial mass function (IMF) of the 1G BHs, following Refs.~\cite{Gerosa:2019zmo,Gerosa:2016vip,Gerosa:2021hsc,Fragione:2020han}, we consider several distributions:  
(i) a uniform distribution for $m_{\rm BH} \in [3,60]\,M_\odot$ (unless otherwise specified); and  
(ii) a power-law distribution $p(m) \propto m^\alpha$ (with $\alpha=-2.4$) restricted to the same mass range.  
For the 1G BHs, we generally do not consider masses beyond $60\,M_\odot$ because pair instability and pulsational pair instability in massive helium cores are expected to suppress the formation of 1G BHs in this mass range~\cite{Belczynski:2016jno}, although the exact value of the upper limit is not known and depends on the details of nuclear burning physics ~\cite{Farmer:2019jed}.

For the initial spin distribution of 1G BHs, we explore two classes of populations: a \emph{high-spin} population, where the spin magnitudes are drawn uniformly as $\chi_{1,2} \in [0,1]$, and a \emph{low-spin} population (following Ref.~\cite{Fuller:2019sxi}), where $\chi_{1,2} \in [0,0.1]$. In addition, we consider Beta distributions(following Ref.~\cite{KAGRA:2021duu}) over the same range of spin magnitudes. Unless otherwise specified, spin-orientation angles are sampled isotropically and motivated by the spherical symmetry of potential host environments such as globular clusters.

We also investigate several binary pairing functions that determine how the component masses $m_{1,2}$ are selected. First, we consider \textit{random pairing}, in which $m_{1}$ and $m_{2}$ are drawn independently from the underlying 1G mass distribution. We then examine two forms of \textit{selective pairing}, in which $m_{1}$ is drawn from the IMF but the distribution of $m_{2}$ is conditioned on $m_{1}$. Following Refs.~\cite{Gerosa:2019zmo,Gerosa:2016vip,Gerosa:2021hsc,Fragione:2020han}, we consider models of the form 
\begin{equation}
p(m_2 | m_1) = m_2^\beta \quad \text{(model~A)}
\label{eq:model_A}
\end{equation}
with $\beta = 6.7$ and for $m_2<m_1$, and models of the form
\begin{equation}
p(m_2 | m_1) = (m_1 + m_2)^\gamma \quad \text{(model~B)}
\label{eq:model_B}
\end{equation}
with $\gamma = 4$. These choices are motivated by the mass-dependence of the cross section for different binary formation channels~\cite{Kritos:2022ggc}.

\subsection{Comparison of retention probabilities}
\label{sec:pret}
First, we investigate how the retention probability changes when using \gwModel{} in place of the older \HLZ{} model.

\begin{figure}
    \centering
    \includegraphics[width=\columnwidth]{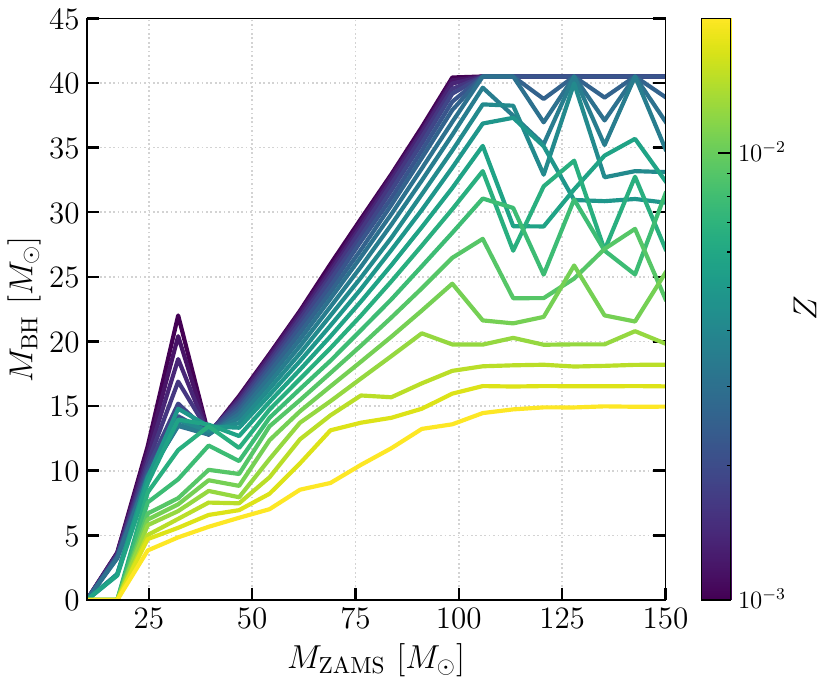}
    \caption{For creating compact objects in our cluster simulations, we use the updated single-star evolution (SSE) delayed prescription of Refs.~\cite{Hurley:2000pk,2002MNRAS.329..897H,Banerjee:2019jjs}. Here, we show the masses of remnant black holes formed from stellar collapse as a function of the progenitor star's ZAMS mass, with metallicity indicated in the color bar. More details are in Section~\ref{sec:pimbh}.}
    \label{fig:MZAMS_MREM}
\end{figure}

In Figure~\ref{fig:pret_q_chimax_grid}, we show the probability (estimated over $10^4$ Monte Carlo realizations to ensure that the results have converged) of retaining the remnant of a BBH merger as a function of the mass ratio $q$ and the maximum spin magnitude $\chi_{\rm max}$, for several cluster escape velocities $v_{\rm esc} = 50$, $100$, $150$, and $200$ km\,s$^{-1}$, using both the \gwModel{} and \HLZ{} prescriptions. We do not vary the individual BH masses here because the recoil kick depends only on the mass ratio and spins, not on the absolute masses. For each $q$, we draw the spin magnitudes $\chi_{1,2}$ uniformly from $[0,\chi_{\rm max}]$, and we sample spin-orientation angles isotropically. The resulting retention probability is shown as a color map.

\begin{figure*}
    \centering
    \includegraphics[width=\textwidth]{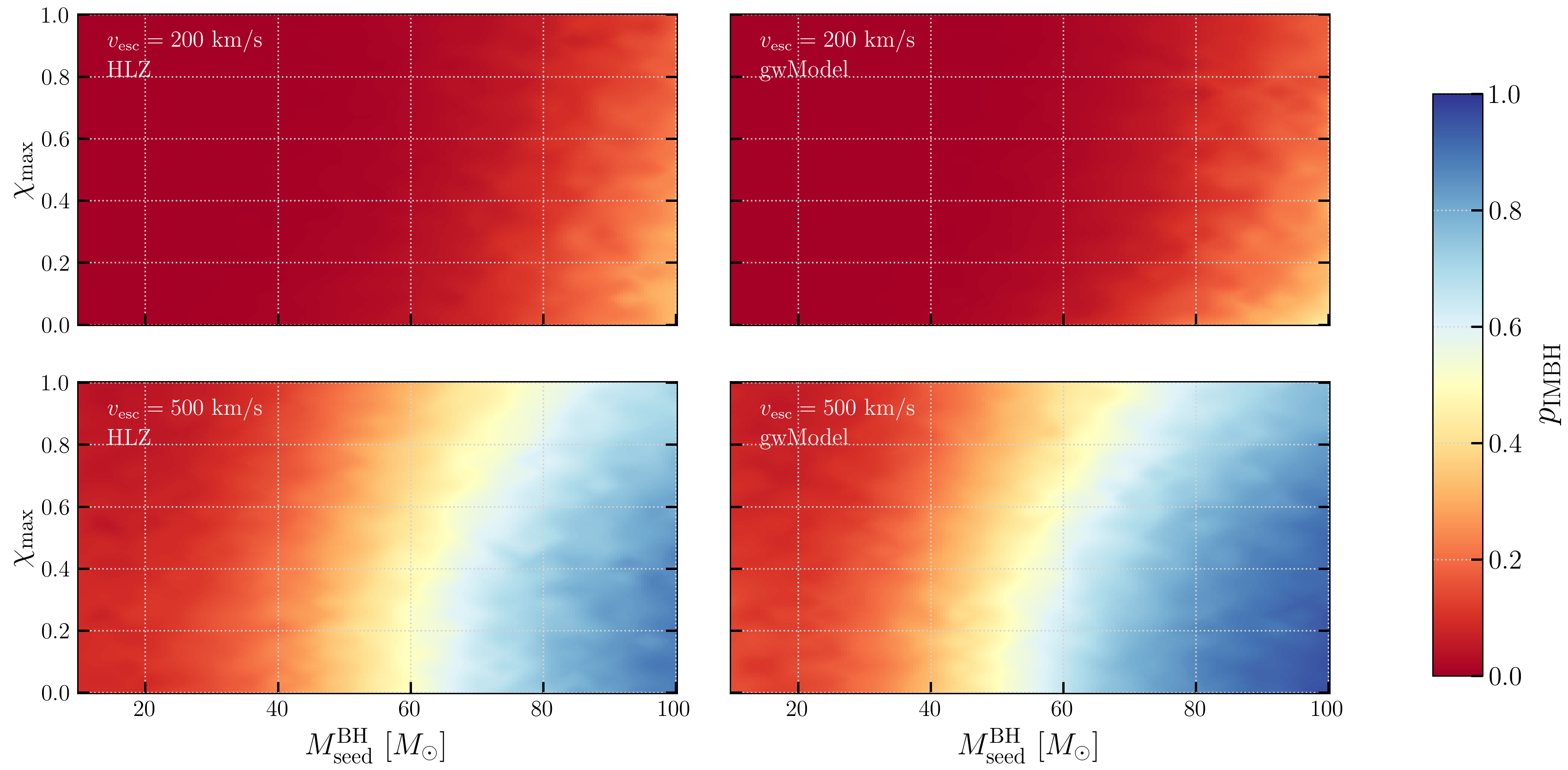}
    \caption{Probability of forming a black hole mass of at least $250\,M_\odot$ through successive mergers as a function of the BH seed mass, shown for different cluster escape speeds: 200 km s$^{-1}$ (top panels), and 500 km s$^{-1}$ (bottom panels). The left columns show the results obtained using the \HLZ{} recoil-kick model, while the right columns show the results for the \gwModel{}. The black-hole spins are drawn from a uniform distribution in the range $[0, \chi_{\max}]$. We choose metalicity of $Z=0.005$. Details are in Section~\ref{sec:pimbh}.}
    \label{fig:pimbh_bhseed_chimax_grid}
\end{figure*}

For $v_{\rm esc} = 50$ km\,s$^{-1}$, we find that almost no remnants are retained when kicks are computed with \HLZ{}. However, when using \gwModel{}, the retention probability increases significantly for higher mass ratios ($q \gtrsim 6$) and for near-equal-mass binaries with $\chi_{\rm max} \lesssim 0.6$. Similar trends appear for escape velocities of 100 and 150 km\,s$^{-1}$. 
For example, at $v_{\rm esc} = 100$ km\,s$^{-1}$, the retention probability increases from nearly zero to values in the range $0.2$--$0.4$ for mass ratios between $q \approx 1.5$ and $q \approx 7$ when switching from \HLZ{} to \gwModel{}. 
For $v_{\rm esc} = 500$ km\,s$^{-1}$, the escape velocity is sufficiently large that nearly all remnant BHs are retained regardless of whether \HLZ{} or \gwModel{} is used, and the differences between the two models largely vanish.

To quantify the differences in retention probabilities predicted by the two models, we compute the JSD between the retention--probability distributions (over all mass ratio and maximum spin values) obtained from \gwModel{} and \HLZ{} at several escape velocities. We find JSD values of 0.351, 0.467, 0.098, 0.100, 0.051, and 0.021 for escape velocities $v_{\rm esc}$ of 25, 50, 100, 150, 200, and 500 km/s, respectively. This reaffirms that observations that the retention probabilities are quite different between \gwModel{} and \HLZ{} for escape velocities less than $500$ km/s. We repeat the same experiment using initial spin magnitudes drawn from a Beta distribution over $[0,\chi_{\rm max}]$. We find qualitatively similar differences between the two models (not shown). 

\subsection{Probability of forming an intermediate-mass black hole}
\label{sec:pimbh}
Next, following Ref.~\cite{Fragione:2020han}, we perform another simplified simulation with minimal assumptions to assess whether hierarchical mergers can lead to the formation of IMBHs with a mass of $100M_{\odot}$ or beyond. In addition, we also ask whether the black hole mass can reach $150M_{\odot}$ (remnant mass of GW190521) and $200M_{\odot}$ (remnant mass of GW231123).

We begin by choosing a metallicity $Z$ for the host cluster, which sets the maximum mass of BHs that can form through stellar collapse. For the stellar population, we consider zero-age main-sequence (ZAMS) masses in the range $[0.08,150]\,M_{\odot}$. We then sample $5000$ stellar masses from a Kroupa mass function.
To compute the mass of the corresponding 1G BHs, we use the updated single-star evolution (SSE) delayed prescription of Refs.~\cite{Hurley:2000pk,2002MNRAS.329..897H,Banerjee:2019jjs}, as implemented in Ref.~\cite{Kritos:2022ggc}.

In Figure~\ref{fig:MZAMS_MREM}, we show the resulting remnant BH masses as a function of the progenitor ZAMS mass and the metallicity. Typically, that at low metallicities (for example, $Z = 0.0002$), stellar collapse can produce more massive BHs, with masses reaching up to $\sim 45\,M_\odot$. As the metallicity increases toward solar values, the maximum BH mass attainable through stellar collapse decreases, with typical upper limits of only $\sim 15\,M_\odot$ in our prescription. Note that this upper limit may vary depending on the specific prescription adopted~\cite{Hurley:2000pk,2002MNRAS.329..897H,Belczynski:2016jno,Farmer:2019jed,Spera:2017fyx}.

As in the previous section, we draw spin magnitudes from $[0,\chi_{\rm max}]$ and sample spin-orientation angles isotropically for the 1G BHs. Additionally, we assume a mass-weighted pairing function $\propto (m_1+m_2)^{4}$.

\begin{figure*}
    \centering
    \includegraphics[width=0.47\textwidth]{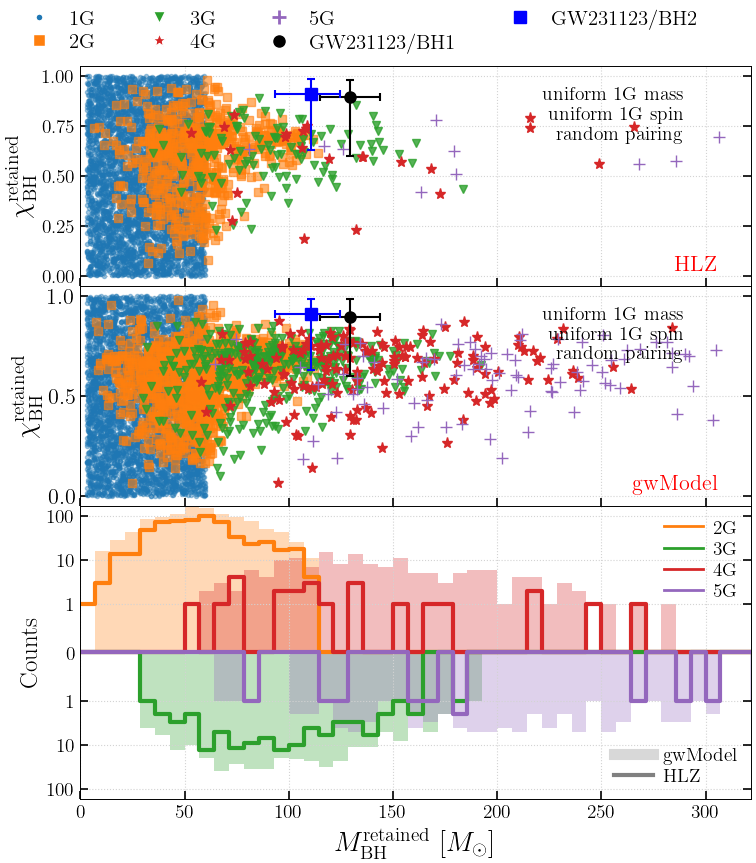}
    \hspace{6mm}
    \includegraphics[width=0.47\textwidth]{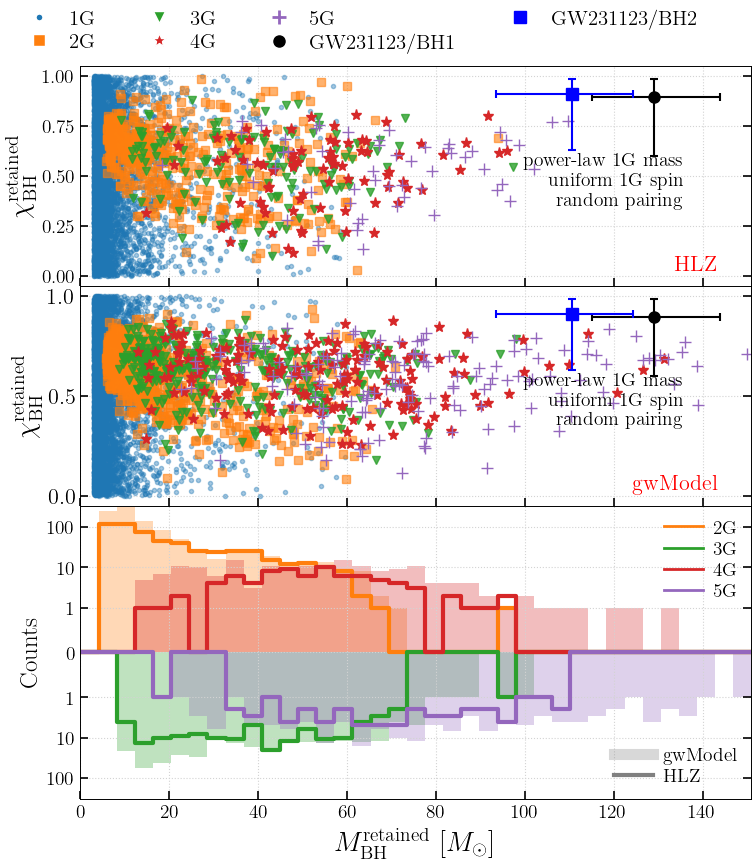}
    \\[1em]
    \includegraphics[width=0.47\textwidth]{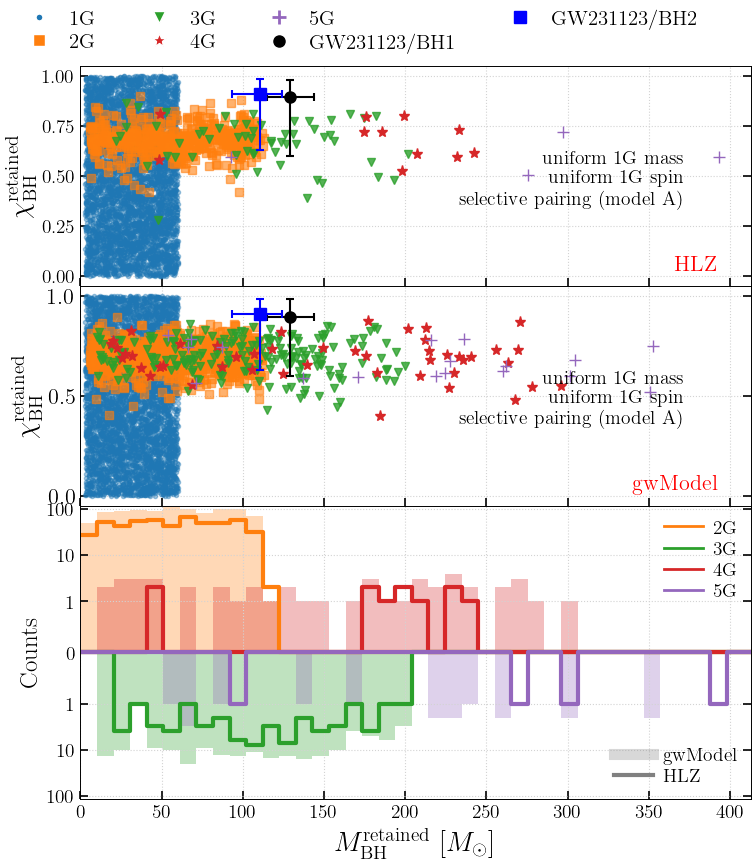}
    \hspace{8mm}
    \includegraphics[width=0.47\textwidth]{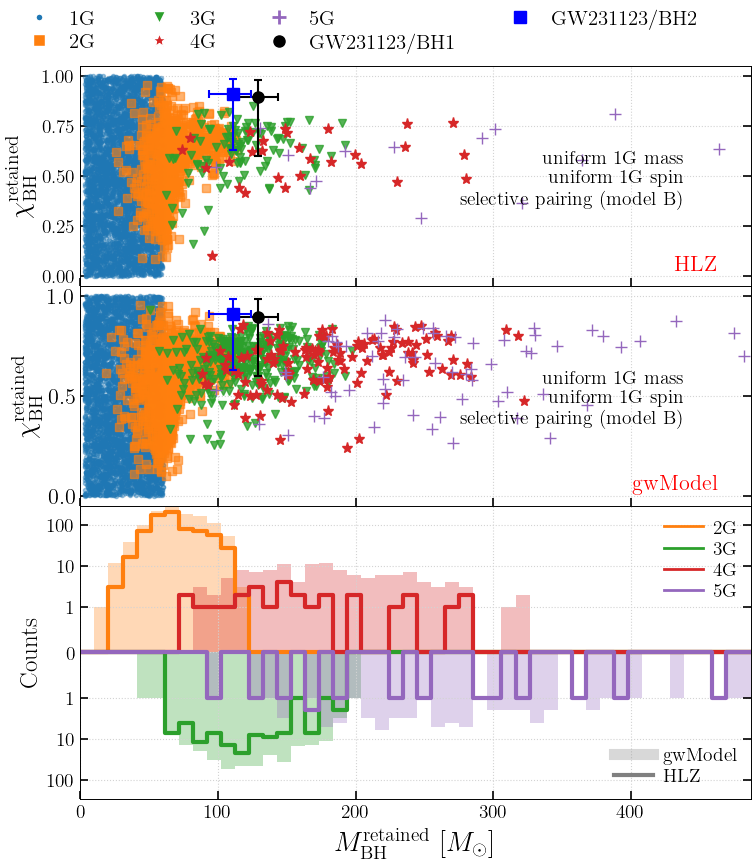}
    \caption{Distribution of retained black-hole masses and spins across five merger generations. For reference, we show the mass and spin inference of BHs in the GW231123 binary in black and blue error bars throughout the paper. In each panel, the upper subplot corresponds to the \HLZ{} model and the middle subplot to the \gwModel{} whereas the lower subplot shows the mass distribution of the retained remnants. The spins of the 1G black-hole population are drawn uniformly from $[0,1]$, and the initial black-hole masses from $[3,60]\,M_\odot$. We vary the initial mass distribution (uniform/power-law) and the pairing function (random/model A/model B, see Eqs.~\ref{eq:model_A} and~\ref{eq:model_B}) one at a time: uniform IMF with random pairing (upper left), power-law IMF with random pairing (upper right), uniform IMF with model A pairing (lower left), and uniform IMF with model B pairing (lower right). In all cases, escape velocities are drawn uniformly from $[1,300]$\,km\,s$^{-1}$, typical of globular clusters. Details are in Section~\ref{sec:toy_ng_problem}.}
    \label{fig:MC_ng_mergers}
\end{figure*}

\begin{figure}
    \includegraphics[width=\columnwidth]{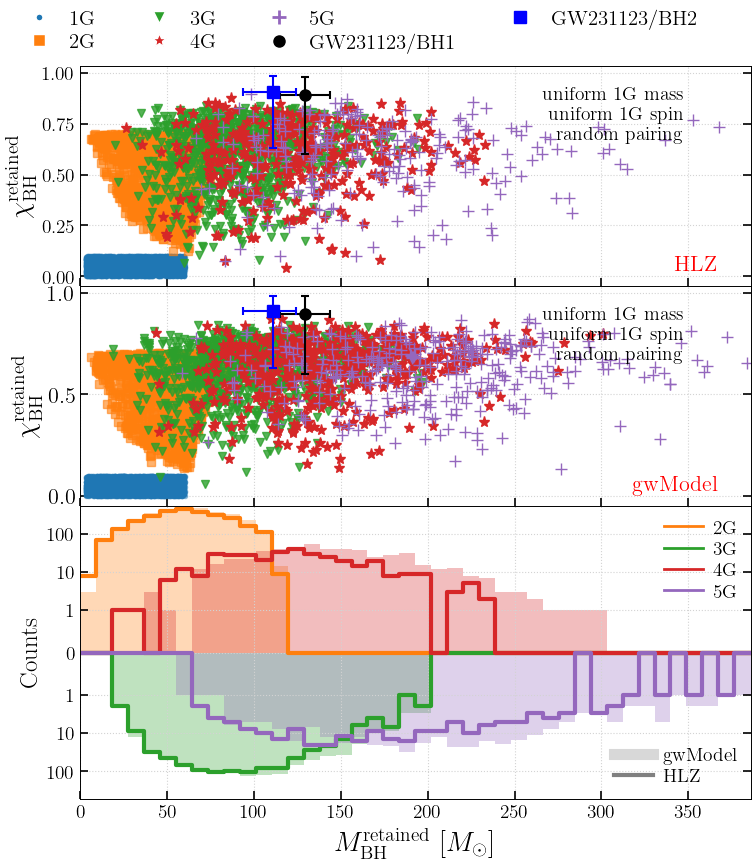}
    \includegraphics[width=\columnwidth]{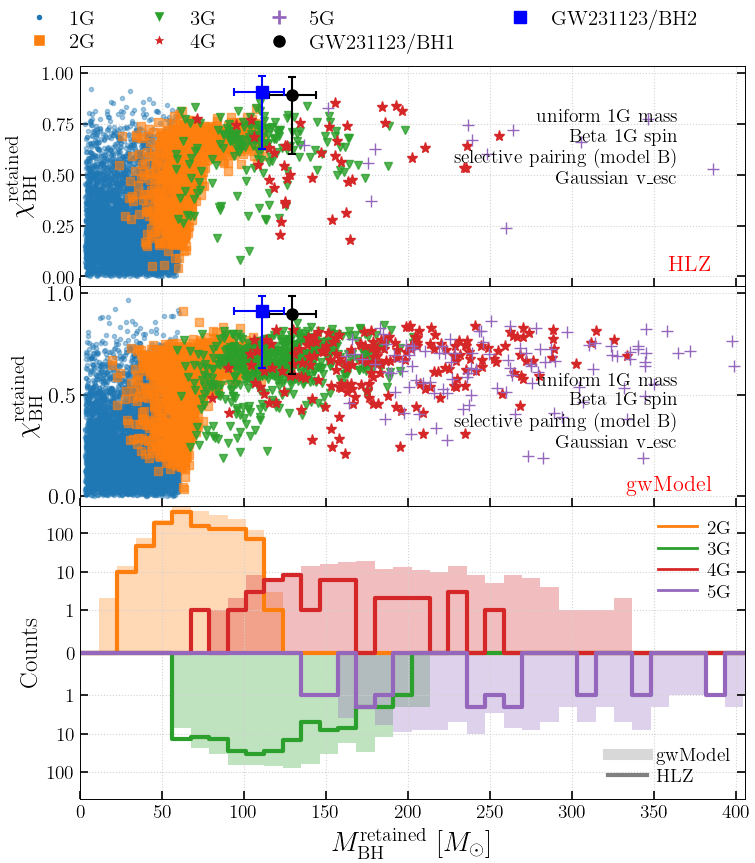}
    \caption{Same as the top-left panel of Fig.~\ref{fig:MC_ng_mergers} but exploring different distributions of 1G BH spins. \textbf{Top:} 1G BH spins drawn uniformly from $[0,0.1]$ instead of $[0,1]$. \textbf{Bottom:} 1G BH spins are drawn from the range $[0,1]$ following a Beta distribution. Additionally, we use the mass-weighted pairing from Eq.~\ref{eq:model_B} when forming binaries, and a Gaussian distribution for escape velocities instead of a uniform distribution. Details are in Section~\ref{sec:toy_ng_problem}.}
    \label{fig:MC_ng_mergers_small_spin}
\end{figure}

On the other hand, collisions or mergers of massive stars may produce BH remnants in the high–mass gap, i.e., beyond $60\,M_\odot$. Motivated by this possibility, we consider an additional population of BHs with masses in the range $[10,100]\,M_\odot$. Following Ref.~\cite{Fragione:2020han}, we treat BHs in this mass range as \emph{seed} BHs. These seed BHs are allowed to merge with the 1G BH population, and any remnant BH that is not ejected from the host environment is permitted to participate in subsequent hierarchical mergers. We then compute the fraction of seed BHs that eventually produce an IMBH of at least $100 M_\odot$ or $250 M_\odot$.
In this experiment, we do not allow the seeds to merge among themselves, nor do we permit the 1G black holes to merge with one another. All mergers occur exclusively between a 1G black hole and a seed (or its remnant). 

To explore the dependence on the host-cluster potential, we consider a range of escape velocities, $v_{\rm esc} \in [50,500]$ km\,s$^{-1}$. Our results are shown in Figure~\ref{fig:pimbh_bhseed_chimax_grid}. For metallicity $Z = 0.005$, we find that for $v_{\rm esc} \lesssim 100$ km\,s$^{-1}$ the probability of forming an IMBH of mass $250M_{\odot}$ is extremely small across all seed masses, since most mergers produce recoil velocities of order $50$–$100$ km\,s$^{-1}$, leading to the ejection of the merger remnants. We therefore omit these cases from the figure.
For escape velocities of $v_{\rm esc} = 200$ km\,s$^{-1}$, we find that only a small fraction of seed BHs reach IMBH masses. On the other hand, for $v_{\rm esc} = 500$ km\,s$^{-1}$, these fraction becomes quite large. Particularly, many black holes with seed masses of $\sim 90\,M_\odot$ and $\sim 60\,M_\odot$ reach IMBH mass of $250\,M_\odot$. Here, we also find that using \gwModel{} leads to a higher fraction of seed BHs forming IMBHs (with mass of $250\,M_\odot$) compared to the \HLZ{} prescription. In most cases, we find that, the maximum mass of the IMBHs formed this way was $\sim 280M_\odot$.

To quantitatively assess how the two recoil--kick prescriptions differ in their predicted IMBH--formation probabilities, we compute the integrated probability difference averaged over the entire parameter space:
\begin{equation}
\Delta P_{\rm IMBH}^{\rm tot} =
\iint \delta P_{\rm IMBH} \, 
{\rm d}M_{\rm seed}\,{\rm d}\chi_{\max},
\end{equation}
where  
\begin{equation}
\delta P_{\rm IMBH} =
P_{\rm IMBH}^{\rm gwModel}(M_{\rm seed}^{\rm BH},\chi_{\max})
-
P_{\rm IMBH}^{\rm HLZ}(M_{\rm seed}^{\rm BH},\chi_{\max}).
\end{equation}
A positive value of $\Delta P_{\rm IMBH}^{\rm tot}$ indicates that \gwModel{} predicts a higher overall probability of IMBH formation.
We find that the cumulative integrated difference is $\Delta P_{\rm IMBH}^{\rm tot} \sim 8\%$ at $v_{\rm esc} = 500~\mathrm{km\,s^{-1}}$.  
For an escape velocity of $50~\mathrm{km\,s^{-1}}$, \gwModel{} predicts a higher probability of forming a $100\,M_\odot$ IMBH in $1\%$ of the parameter space compared to \HLZ{}.  
The corresponding fractions are $20.8\%$, $31.7\%$, and $100\%$ for escape velocities of $150$, $200$, and $500~\mathrm{km\,s^{-1}}$, respectively. 
This qualitative behavior remains the same when comparing the probabilities of forming IMBHs with masses of $150\,M_\odot$ and $250\,M_\odot$. 
We find that these results are not strongly dependent on the assumed maximum mass of the seed black holes (not shown). More analyses are presented in Appendix~\ref{sec:appendix}. 

\subsection{Comparison of the mass and spin distributions across merger generations}
\label{sec:toy_ng_problem}
Our next set of experiments tracks the masses and spins of retained BHs across successive merger generations in simplified cluster environments. In hierarchical merger scenarios, we define the BH generation as follows. BHs formed through direct stellar collapse are denoted as 1G. Black holes formed via the merger of two 1G BHs are classified as second-generation (2G). Black holes formed through the merger of a 2G BH with either a 1G or another 2G BH are classified as third-generation (3G), and so on.

To mimic the diversity of cluster potentials, we draw the escape velocity uniformly from the range $[1,300]$ km\,s$^{-1}$. We begin with $5000$ 1G BHs whose masses and spins are sampled from specified distributions. We then use a pairing function among the options listed below, and each resulting binary is merged to produce a population of 2G BHs. The remnant masses and spins are computed using semi-analytical formulae calibrated to NR data~\cite{Barausse:2012qz, Barausse:2009uz}. These fits are known to be highly accurate, with modeling errors typically below $1\%$, and can therefore be treated as sufficiently precise for astrophysical applications. For the recoil kick, we again consider both the \gwModel{} and \HLZ{} prescriptions.
For each 2G BH, we compare the recoil kick velocity to the assigned escape velocity and thus decide if the binary is retained in the cluster. We then allow the retained 2G BHs to merge among themselves as well as with 1G BHs, producing a population of 3G BHs. This procedure is repeated iteratively, and we track the growth of BHs up to 5G. We perform these experiments for a variety of initial mass and spin distributions, as well as different binary pairing functions.


In Figure~\ref{fig:MC_ng_mergers}, we show the distributions of retained BH masses and spins across five merger generations. For reference, we also display the mass and spin inference for GW231123. In each panel, the upper subplot corresponds to the \HLZ{} model, while the lower subplot corresponds to the \gwModel{}. The spins of the 1G black-hole population are drawn uniformly from $[0,1]$, and the initial black-hole masses are in the range $[3,60]\,M_\odot$.

For the initial BH masses, we show results for two different distributions: uniform and power-law.
We also consider three pairing functions for assembling binaries. Our default choice is random mass pairing, and we additionally use two selective pairing models introduced earlier in this section: model~A \cite{Gerosa:2019zmo}, and model~B~\cite{Fragione:2020han,OLeary:2016ayz} (see Eqs.~\ref{eq:model_A} and~\ref{eq:model_B}). Finally, escape velocities are drawn uniformly from $[1,300]\,\mathrm{km\,s^{-1}}$, typical of globular clusters. We vary the initial mass function and pairing function one at a time in the different panels of the figure.

Our main findings from Fig.~\ref{fig:MC_ng_mergers} are as follows:  
(1) The distributions of BH masses and spins across successive generations change noticeably when replacing the \HLZ{} kick prescription with the more accurate \gwModel{}, regardless of the assumed initial mass or pairing function.
(2) The \gwModel{} consistently predicts a larger number of BHs occupying the mass–spin region consistent with the inferred properties of GW231123.  
(3) More massive BHs are formed when the initial mass function follows a Kroupa distribution.
(4) We find that pairing model~A generally produces fewer massive BHs across successive merger generations, whereas the pairing model~B yields a noticeably larger population of high-mass BHs. 
(5) Finally, when the 1G BH population is drawn from a uniform mass distribution instead of a power-law distribution, we find that more massive black holes are produced through hierarchical mergers.

Next, we show the impact of the initial spin distribution in Figure~\ref{fig:MC_ng_mergers_small_spin}. The top panel shows results for 1G BH spins drawn uniformly in a narrow range $[0,0.1]$, while assuming a uniform initial mass function and random pairing. We find that slowly spinning initial BHs produce more massive merger remnants than an initial spin distribution in $[0,1]$ (cf. Figure~\ref{fig:MC_ng_mergers}; upper left). The bottom panel of Fig.~\ref{fig:MC_ng_mergers_small_spin} shows the impact of varying several ingredients simultaneously. We show the distributions of retained BH masses and spins across five merger generations for a setup in which the initial BH spin magnitudes are drawn from a Beta distribution over $[0,1]$, the initial BH masses are drawn uniformly from $[3,60]\,M_\odot$, and a selective, mass-weighted pairing function is used when forming binaries. In addition, escape velocities are sampled from a Gaussian distribution spanning $[1,300]$ km\,s$^{-1}$. 

We also have experimented with other combinations of initial mass functions (such as power-law) and pairing functions (such as model A and model B). As in previous cases, we find that the \gwModel{} prescription produces more massive BHs than the \HLZ{} model.

\subsection{Comparison of the retention probability across cluster properties}
\label{sec:pret_simple_cluster}
Our next experiment incorporates more realistic astrophysical information about properties of globular clusters into the model. We simulate 5000 BBH mergers in each of 1500 clusters, varying the metallicity, cluster mass, and half-mass radius. The metallicity $Z$ is drawn uniformly in logarithmic space between $0.0002$ and $0.02$. Cluster masses $M_{\rm cl}$ are sampled uniformly in logarithmic space over the range $10^{4}$--$10^{7}\,M_\odot$, and the half-mass radius $r_{\rm h}$ is selected uniformly in logarithmic space between $0.1$ and $10$ pc. 

\begin{figure}
    \centering
    \includegraphics[width=\columnwidth]{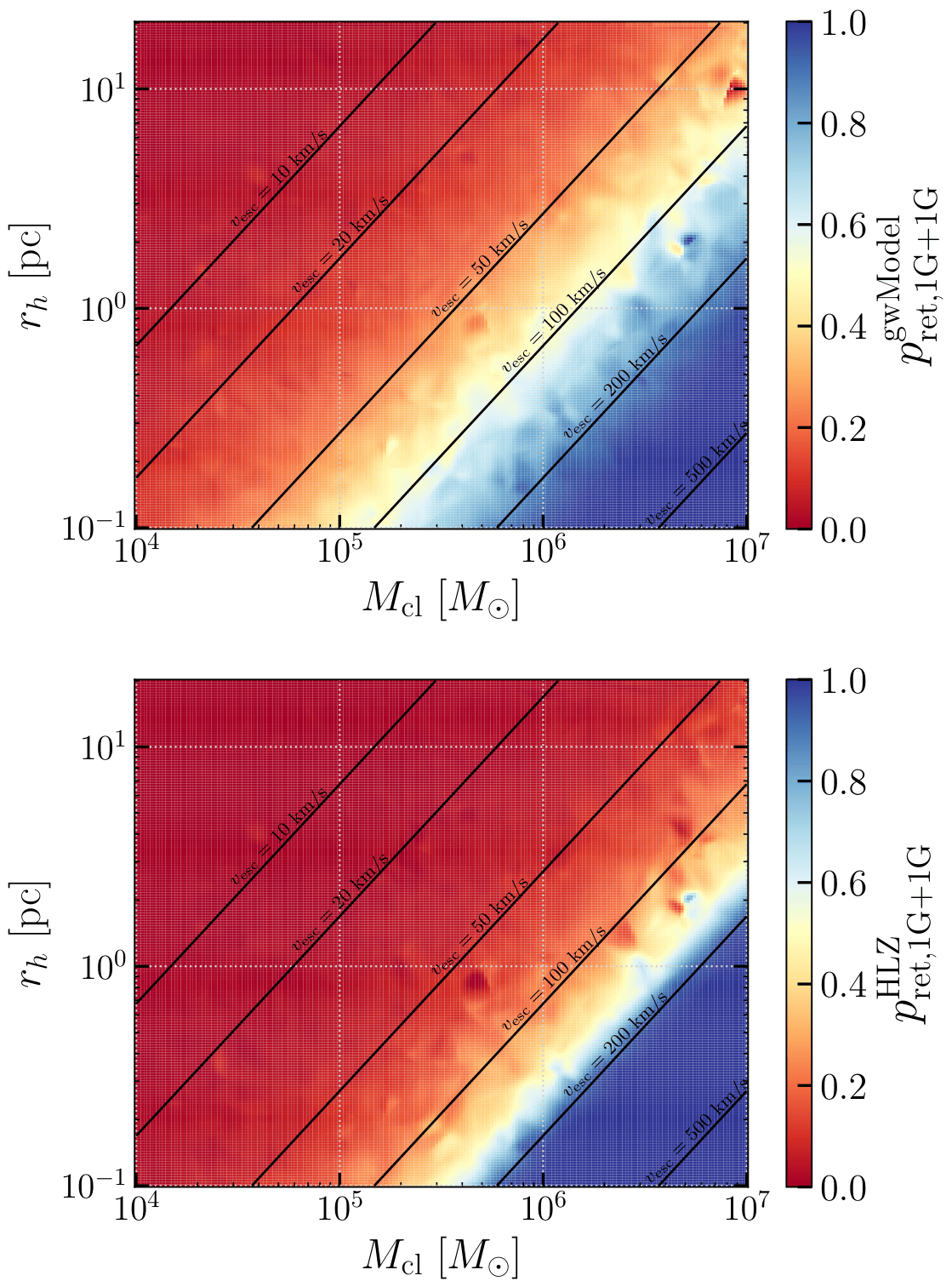}
    \caption{In Figs.~\ref{fig:MC_ng_mergers} and~\ref{fig:MC_ng_mergers_small_spin}, we had marginalized over cluster properties (using the prior for cluster escape velocities as uniform over $[1,300]$\,km\,s$^{-1}$). Here, we show the variation of the probability of retaining the merger remnant of two 1G BHs with cluster mass and half-mass radius (for reference, the lines of constant escape velocity are in black). The top (bottom) panel shows results obtained using the \gwModel{} (\HLZ{}) recoil-kick prescription. The largest differences in the kick models are observed in clusters with $v_{\rm esc}$ in the range 30--300 km/s. Other population properties for BH mass and spin distributions are similar to the top-right panel of Fig.~\ref{fig:MC_ng_mergers}, see Section~\ref{sec:pret_simple_cluster} for further details.
    }
    \label{fig:simple_cluster_retention}
\end{figure}

For the 1G BHs, the spin magnitudes $\chi_{1,2}$ are drawn uniformly from $[0,1]$, and all 1G BHs are assumed to form via stellar collapse. The ZAMS masses of the progenitor stars, $M_{\rm ZAMS}$, are sampled uniformly from the range $[10,150]\,M_\odot$. The cluster metallicity determines the resulting distribution of 1G BH masses. The cluster mass and half-mass radius determine the cluster escape velocity, $v_{\rm esc}$~\cite{Kritos:2022ggc}:
\begin{equation}
v_{\rm esc}
\simeq 26~{\rm km\,s^{-1}}
\left( \frac{M_{\rm cl}}{10^5\,M_\odot} \right)^{1/2}
\left( \frac{r_h}{1~{\rm pc}} \right)^{-1/2}
\end{equation}
As in previous sections, spin-orientation angles are drawn isotropically. We assume random mass pairing between the two BHs forming a binary. We then compute the retention probability of the 1G+1G merger remnants.


\begin{figure*}
    \centering
    \includegraphics[width=0.98\textwidth]{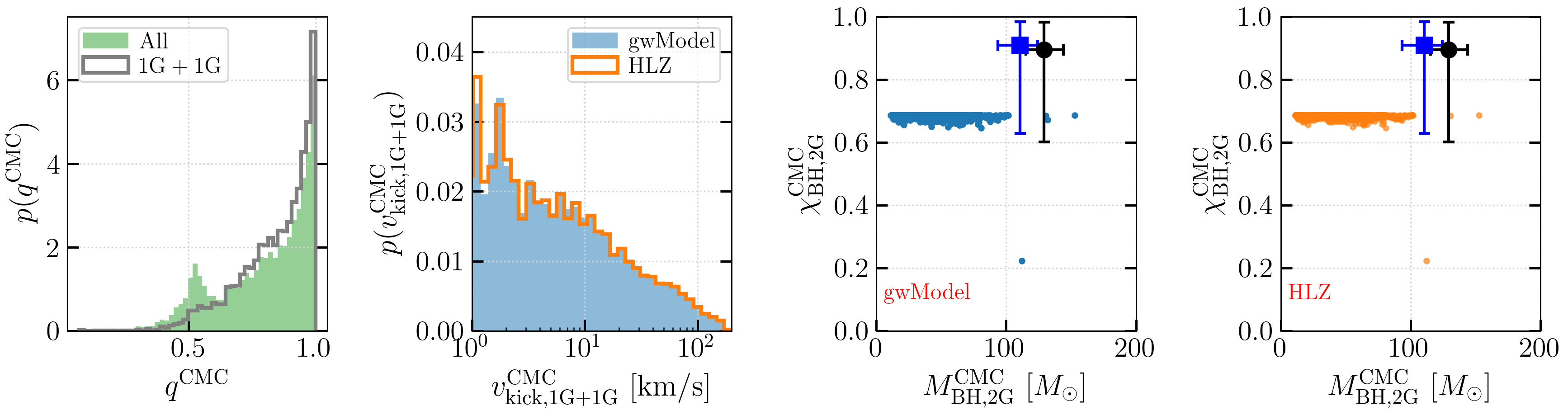}\\
    \includegraphics[width=0.98\textwidth]{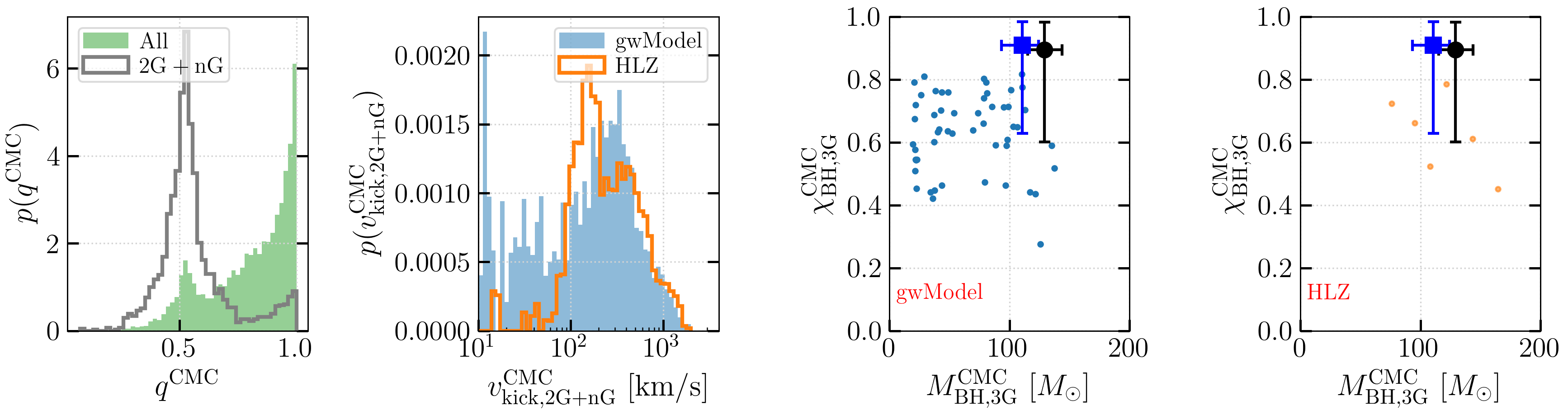}\\
    \includegraphics[width=0.98\textwidth]{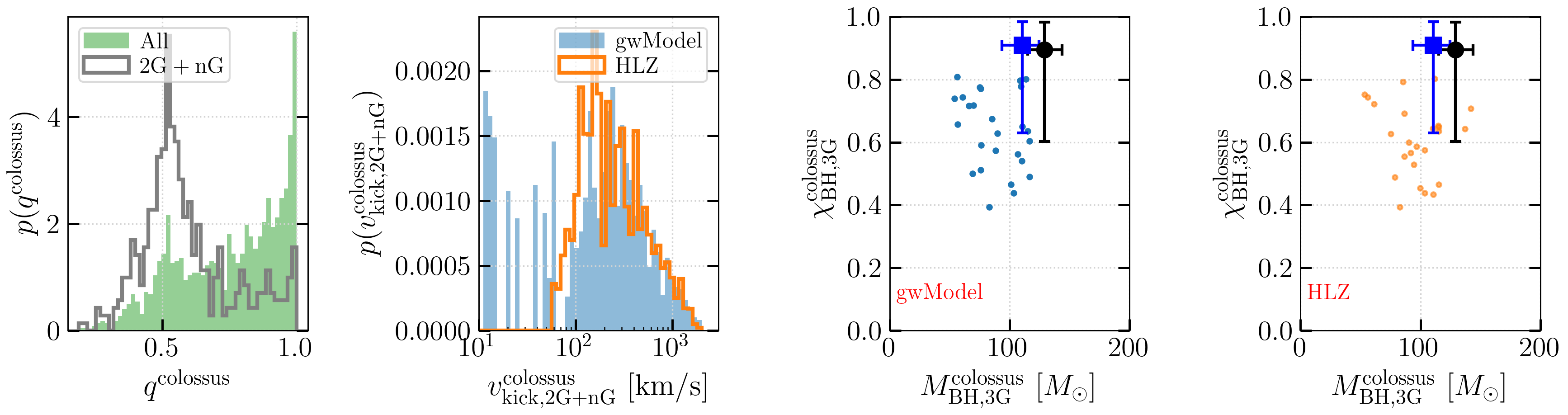}\\
    \caption{We now test the two recoil-kick models using detailed cluster-evolution simulations (\textcolor{linkcolor}{\texttt{CMC}}, \textcolor{linkcolor}{\texttt{colossus}}) that self-consistently model the cluster evolution and BH mergers. \textbf{\textit{Upper panels:}} The leftmost panel shows the mass-ratio distribution of all mergers (green) and 1G+1G mergers only (gray), obtained from the \textcolor{linkcolor}{\texttt{CMC}} catalog. The second panel shows the recoil kick velocities for all 1G+1G mergers. The third and fourth panels show the mass and spin distributions of the \emph{retained remnant BHs} after 1G+1G mergers. We find that, because 1G BHs are non-spinning, both recoil prescriptions yield similar kick velocities and produce similar mass and spin distributions for the retained 2G BHs. 
    We show analogous results for 2G+1G mergers of the \textcolor{linkcolor}{\texttt{CMC}} catalog (\textbf{\textit{center panels}}), where the mass ratios are more asymmetric and the BH spins are non-zero. The differences between the \gwModel{} and \HLZ{} recoil velocities are less pronounced compared to previous Figs.~\ref{fig:MC_ng_mergers} and~\ref{fig:MC_ng_mergers_small_spin}, but they are still noteworthy. The models also yield slightly different mass and spin distributions for the retained 3G BHs.
    Finally, we show the analogous results for 2G+1G mergers from the \textcolor{linkcolor}{\texttt{colossus}} (\textbf{\textit{bottom panels}}) catalog. 
     For reference, we also show the inferred mass and spin of the two BHs involved in GW231123 (as black and blue markers with error-bars).  Details are in Section~\ref{sec:cmcpostprocess}.}
    \label{fig:cmc_colossus_rapster}
\end{figure*}

In Figure~\ref{fig:simple_cluster_retention}, we show the retention probability as a function of cluster mass $(M_{\rm cl}$ and half-mass radius $r_{\rm h})$. We observe lines of nearly constant escape velocity that govern the retention probability. As expected, more massive clusters with smaller half-mass radii tend to retain a larger fraction of 1G+1G merger remnants. Furthermore, when using \gwModel{}, a larger fraction of remnants are retained compared to the \HLZ{} prescription, consistent with the fact that \HLZ{} generally predicts higher kick velocities. In particular, for clusters with escape velocities in the range $20\,\mathrm{km\,s^{-1}}$ to $200\,\mathrm{km\,s^{-1}}$, the differences in retention probabilities predicted by \gwModel{} and \HLZ{} become more pronounced.
Additionally, we find that cluster metallicity has a little effect on the retention probability (not shown in the figure). This is not unexpected: although metallicity influences the masses of the 1G BHs, the recoil kick depends primarily on the mass ratio. Under random pairing, the mass-ratio distribution is only weakly sensitive to the underlying BH mass distribution, leading to similar retention probabilities across metallicities.

\section{Detailed cluster simulations}
\label{sec:rapster}

We now move to more detailed cluster-evolution simulations that self-consistently follow the collisional dynamics, stellar and binary evolution, dynamical interactions, hierarchical black hole mergers and other relevant astrophysical processes. A key distinction between the analytic estimates of Section~\ref{sec:simple_calculations} and the $N$-body/semi-analytical simulations presented here is that the latter self-consistently evolve the cluster over time. As binaries harden in the cluster core, energy conservation requires that the liberated binding energy be deposited into the surrounding stellar 
population. This drives cluster expansion via Hénon's principle and causes the half-mass radius to grow secularly. Simultaneously, BH ejections through dynamical recoil reduce the total cluster mass. Together, these two effects cause the cluster escape velocity, $v_\mathrm{esc} \propto \sqrt{M_\mathrm{cl}/r_h}$, to decrease monotonically over time.

\begin{figure}
    \centering
    \includegraphics[width=\columnwidth]{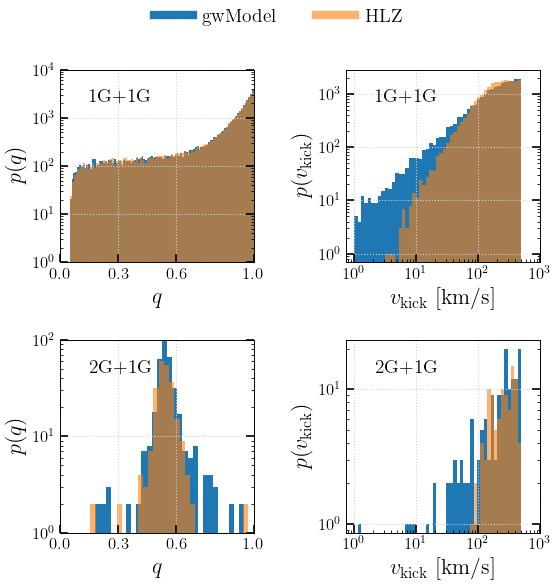}
    \caption{Similar to the left two columns of Fig.~\ref{fig:cmc_colossus_rapster} but for semi-analytical code \rapster{}. Here, we do not show results for an ensemble of clusters, but instead for a representative cluster with different random seeds for sampling the BH masses and spins magnitude and direction.
    The cluster is initiated with $N = 2.5 \times 10^{6}$ stars drawn at metallicity $Z = 0.002$, and with a half-mass radius of $r_{\rm h} = 1\,\mathrm{pc}$ (i.e. with initial escape velocity of $\sim 118$ km/s). Details are in Section~\ref{sec:rapster_new}.}
    \label{fig:kick_distribution_cluster_4}
\end{figure}

\begin{figure*}
    \centering
    \includegraphics[width=\textwidth]{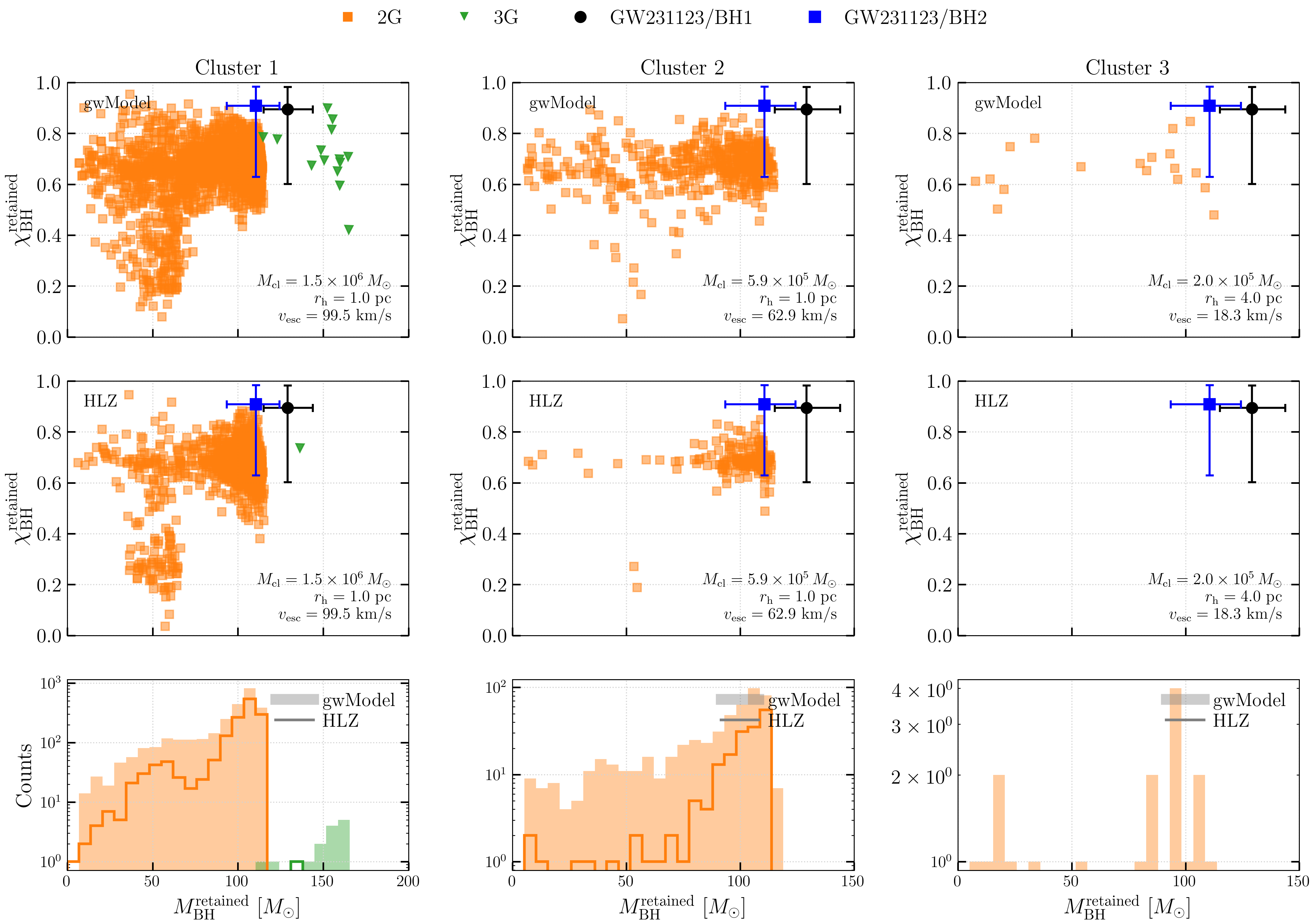}
    \caption{Similar to Fig.~\ref{fig:kick_distribution_cluster_4} but showing the mass and spin distributions of retained BHs in different \rapster{} simulations from the two recoil kick models. We show results from three distinct clusters (each with $2500$ realizations).
    The clusters have initial escape velocities of $\sim 99.5$ km/s, $\sim 62.9$ km/s and $\sim 18.3$ km/s respectively and were chosen such that their final mass and radius ($M_\mathrm{cl}\sim 2\times 10^5 M_\odot$ and $R_\mathrm{cl}\sim 4$ pc) is close to the median of the Galactic Globular Cluster Database$^{\ref{foot:gc_database}}$, Version 4 \cite{Baumgardt18_GC_catalog}. 
    The lowest panel shows the 1D histogram of the retained remnant masses. For reference, we again  show the inferred mass and spin of the two BHs involved in GW231123.
    Details are in Section~\ref{sec:rapster_new}.}
    \label{fig:rapster}
\end{figure*}

\subsection{Post-processing existing cluster simulations}
\label{sec:cmcpostprocess}
We begin by post-processing existing indirect N-body and semi-analytical star cluster simulations spanning a grid of initial cluster properties, including the number of stars, half-mass radius, and metallicity. These simulations are routinely used to understand the binary properties of GW sources and to compare with detected events. They therefore provide an ideal and well-tested framework for our investigation. 
Here, we use only publicly available cluster evolution data from the \textcolor{linkcolor}{\texttt{CMC}}~\cite{Kremer:2019iul}~\footnote{\href{https://cmc.ciera.northwestern.edu/BBHmergers_CMCcatalog.dat}{https://cmc.ciera.northwestern.edu/BBHmergers\_CMCcatalog.dat}} and \textcolor{linkcolor}{\texttt{colossus}}~\cite{Mai:2025jmk}~\footnote{\href{https://cmc.ciera.northwestern.edu/BBHmergers_colossus.dat}{https://cmc.ciera.northwestern.edu/BBHmergers\_colossus.dat}} catalogs.
These frameworks model stellar evolution, relaxation-driven structural changes, and the dynamical interactions that lead to BH mergers. 
Here, we only mention the physics included in the simulation without elaborating the details of their implementation.  

Each cluster simulation starts with $N$ stars drawn from a chosen Kroupa initial mass function at a chosen metallicity, with a half-mass radius of $r_{\rm h}$. This translates to an initial total cluster mass. The evolution framework then maps the ZAMS mass distribution to compact remnants using precomputed stellar-evolution prescriptions, and computes the initial cluster mass, velocity dispersion, and escape velocity under the assumption of virial equilibrium. The cluster subsequently evolves through stellar-evolution mass loss, two-body relaxation, and evaporation, all of which reduce $M_{\rm cl}(t)$ and $v_{\rm esc}(t)$ while driving an expansion of $r_{\rm h}(t)$. The time dependence of the escape velocity is particularly important for determining whether BH merger remnants remain bound to the cluster.

Within this evolving background, BH binaries form and merge through multiple dynamical channels, including three-body binary formation, binary--single and binary--binary encounters and GW captures in close passages.These frameworks further employs mass-weighted pairing, which preferentially selects the most massive BHs when forming binaries and therefore facilitates hierarchical growth when remnants are retained. Both for two-body and three-body dynamical interactions, the BH masses are chosen using a mass-weighted distributions. 

For each BH merger event, the code computes the remnant mass, spin, and GW recoil kick. For the kick, the default choice is to use \HLZ{} prescription. A remnant is retained if its recoil velocity is smaller than $v_\mathrm{esc}$ at the time of merger; otherwise, it is ejected and removed from further evolution. This setup enables us to track the formation pathways and hierarchical assembly of BHs within a self-consistently evolving cluster environment.

Another important piece of physics included in these simulations is the exchange mechanism in three-body interactions, in which the lighter BH in a binary is frequently exchanged with the incoming third BH. This process, often referred to as dynamical hardening, tends to produce more equal-mass binaries in the cluster.

The \textcolor{linkcolor}{\texttt{CMC}} cluster simulations span a $4\times4\times3\times3$ grid in the following parameters: the number of particles $N=\{2\times10^{5},\,4\times10^{5},\,8\times10^{5},\,1.6\times10^{6}\}$, the initial virial radius $r_{\rm v}=\{0.5,\,1,\,2,\,4\}\,\mathrm{pc}$, the metallicity $Z/Z_{\odot}=\{0.01,\,0.1,\,1\}$, and the galactocentric distance $R_{\rm gc}=\{2,\,8,\,20\}\,\mathrm{kpc}$. These parameters are chosen to broadly represent globular clusters in the Milky Way. Here, $N$ denotes the initial number of stars, $r_{\rm v}$ is the virial radius which is related to the half-mass radius through $r_{\rm h}=0.8\,r_{\rm v}$, and $R_{\rm gc}$ is the galactocentric distance.
On the other hand, the \textcolor{linkcolor}{\texttt{colossus}} cluster simulations assume an initial virial radius $r_v = 2\,\mathrm{pc}$, an initial metallicity $Z = 0.1\,Z_{\odot}$, and a galactocentric radius $R_{\rm gc} = 20\,\mathrm{kpc}$ in a Milky Way--like environment, with the number of particles set to $N = 10^7$.
This implies that the escape velocities of clusters in the \textcolor{linkcolor}{\texttt{CMC}} simulations lie in the range $0$--$150\,\mathrm{km\,s^{-1}}$, while the escape velocities in the \textcolor{linkcolor}{\texttt{colossus}} cluster simulations can reach up to $180\,\mathrm{km\,s^{-1}}$.

We first post-process the BBH mergers from the \textcolor{linkcolor}{\texttt{CMC}} catalog. In the left-most panel of Figure~\ref{fig:cmc_colossus_rapster}, we compare the mass-ratio distribution of all mergers with that of the 1G+1G mergers. We find a large abundance of nearly equal-mass BBHs Naturally, we find a large peak at $q \sim 1$ in the 1G+1G case. All 1G BHs are assumed to be non-spinning in the \textcolor{linkcolor}{\texttt{CMC}} catalog. Using this population, we compute the recoil kick velocities for all 1G+1G mergers with the \gwModel{} and \HLZ{} prescriptions, finding that both models predict very similar kick velocities. Consequently, they yield comparable mass and spin distributions for the retained 2G BHs.

Because the underlying mass and spin distributions of 1G and 2G BHs are nearly identical for both recoil prescriptions, we consistently post-process the subsequent 2G+1G mergers using both models. We focus on 2G+1G, as it produces the largest number of mergers after the 1G+1G mergers. The mass-ratio distribution for 2G+1G mergers peaks around $q \sim 0.5$. In this regime, the \gwModel{} and \HLZ{} prescriptions predict noticeably different recoil kick velocities, leading to distinct mass and spin distributions for the retained 3G BHs. For reference, we also show the inferred masses and spins of the two black holes involved in GW231123.

Next, we shift our focus to the \textcolor{linkcolor}{\texttt{colossus}} catalog. In both cases, the initial BHs are also assumed to be non-spinning. Furthermore, the \rapster{} simulations are designed to allow direct comparison with the \textcolor{linkcolor}{\texttt{CMC}} catalog. As before, we find no significant differences in the recoil kick velocities for 1G+1G mergers, nor in the resulting mass and spin distributions of the retained 2G BHs. However, the recoil kick velocities for 2G+1G mergers differ noticeably between the \gwModel{} and \HLZ{} prescriptions. As a result, these differences lead to slightly different mass and spin distributions for the retained 3G BHs.

\subsection{New semi-analytical cluster simulations with \rapster{}}
\label{sec:rapster_new}
Motivated by the results above, We perform new semi-analytical cluster simulations using \rapster{}. We adopt \rapster{} over the \textcolor{linkcolor}{\texttt{CMC}} code primarily for computational 
efficiency: \rapster{} simulates a full globular cluster evolution on a single 
CPU in under a minute, orders of magnitude faster than Monte Carlo 
$N$-body codes. This speedup arises because \rapster{} foregoes individual 
particle tracking entirely, instead evolving a small set of global cluster 
properties (core density, velocity dispersion, and BH population) and 
treating each dynamical channel (three-body binary formation, exchange 
encounters, GW captures) as a Poisson process with analytically prescribed 
rates. This makes \rapster{} well-suited for population-level studies 
requiring large number of cluster simulations.
One other difference between \rapster{} and \textcolor{linkcolor}{\texttt{CMC}} 
is that \rapster{} explicitly tracks mergers from long-lived dynamically assembled 
triples undergoing Lidov--Kozai oscillations as a distinct formation channel, 
whereas \texttt{CMC} only encounters such configurations transiently during 
few-body scattering integrations. For a detailed description 
of the code and its underlying prescriptions, we refer the reader to 
\citet{Kritos:2022ggc}.

We simulate BH mergers in a three representative clusters with different cluster mass $M_{\rm cl}$ and half-mass radius $r_h$ (and therefore different initial escape velocities) using \rapster{}.
Rather than simulating a large ensemble of clusters with different initial conditions, we choose to simulate three representative clusters using $2500$ independent random seeds. This allows us to isolate the impact of different recoil-kick prescriptions on the astrophysical interpretation of hierarchical BH mergers. The representative clusters were chosen such that their final mass and radius is close to the median of the Galactic Globular Cluster Database\footnote{\url{https://people.smp.uq.edu.au/HolgerBaumgardt/globular/parameter.html}\label{foot:gc_database}}, Version 4 \cite{Baumgardt18_GC_catalog}.

The first representative cluster has $N = 2.5 \times 10^{6}$ stars drawn at metallicity $Z = 0.002$ (roughly 10\% of the solar metallicity), and with a half-mass radius of $r_{\rm h} = 1\,\mathrm{pc}$. This corresponds to an initial total cluster mass of $\sim 1.5 \times 10^{6}\,M_{\odot}$ since the initial average stellar mass is $\simeq0.6\,M_\odot$. We also adopt slightly different assumptions for the initial BH populations and binary interactions compared to the previous subsection. Such modifications are straightforward to implement in semi-analytical frameworks like \rapster{} as it enables thousands of simulations to be performed within a relatively short time. We simulate clusters in which the initial BH masses are drawn uniformly from $[3,60]\,M_{\odot}$, while the BH spins are drawn uniformly from $[0,1]$. We adopt random mass pairing in both two-body and three-body interactions. All other aspects of the binary hardening prescription are kept unchanged from the default implementation in \rapster{}.
For this cluster, we find an initial escape velocity of $v_{\rm esc}\sim 100\,\mathrm{km\,s^{-1}}$. 
We first examine the mass-ratio and recoil kick velocity distributions of all mergers (Figure~\ref{fig:kick_distribution_cluster_4}). For the 1G+1G population, we find that the kick distributions predicted by the two models differ noticeably: \gwModel{} yields a larger fraction of low-velocity kicks compared to \HLZ{}. We find that $40{,}616$ mergers occur in-cluster when \gwModel{} is used, compared to $40{,}765$ when \HLZ{} is employed. For \gwModel{} (\HLZ{}), approximately $6\%$ ($3\%$) of 1G+1G mergers produce recoil velocities below $50\,\mathrm{km\,s^{-1}}$. 
These fractions increase to $13\%$ ($9\%$), $27\%$ ($26\%$), and $57\%$ ($57\%$) for recoil thresholds of $100\,\mathrm{km\,s^{-1}}$, $150\,\mathrm{km\,s^{-1}}$, and $500\,\mathrm{km\,s^{-1}}$, respectively. 
Overall, $7\%$ ($3.7\%$) of all mergers yield recoil velocities below the escape velocity of $99.5\,\mathrm{km\,s^{-1}}$ when using \gwModel{} (\HLZ{}).
This means that they retain slightly different number of black holes in the next generation. 
For 1G+2G mergers, these differences within the escape velocity become more pronounced. For higher-generation mergers, the number of mergers is comparatively small, but we still observe a systematic trend in which \gwModel{} predicts smaller recoil velocities than \HLZ{}. For example, we record 372 in-cluster 2G+1G mergers when using \gwModel{}, whereas \HLZ{} yields 241 such mergers. We find that \gwModel{} predicts that approximately $4\%$ of these mergers have recoil velocities below $50\,\mathrm{km\,s^{-1}}$, while \HLZ{} yields none in this range. This behavior is consistent with the increasing divergence between the models at larger mass ratios, where \HLZ{} exhibits more extended high-velocity tails.
In Figure~\ref{fig:rapster} (first column), we show the mass and spin distributions of retained BHs across different merger generations using two recoil-kick prescriptions: \gwModel{} and \HLZ{}. The distributions of retained 2G and 3G BHs differ substantially. We find that none of the models predict a population of retained 4G BHs.

Next, we simulate a slightly less massive cluster, initialized with $N = 10^{6}$ stars at metallicity $Z = 0.002$ and with a half-mass radius of $r_{\rm h} = 1\,\mathrm{pc}$. This corresponds to an initial total cluster mass of $\sim 5.9 \times 10^{5}\,M_{\odot}$ and an initial escape velocity of $v_{\rm esc} \sim 63\,\mathrm{km\,s^{-1}}$. We adopt the same initial mass and spin distributions for the BHs as in the previous case. As before, we find significant differences in the properties of retained 2G BHs between the two recoil prescriptions. In addition, we find that none of the 2G+$n$G merger remnants are retained in this cluster.

Finally, we simulate a light cluster, initialized with $N = 3.4 \times 10^{5}$ stars at metallicity $Z = 0.002$ and with a half-mass radius of $r_{\rm h} = 4\,\mathrm{pc}$ (Figure~\ref{fig:rapster}; third column). This corresponds to an initial total cluster mass of $\sim 2 \times 10^{5}\,M_{\odot}$ and an initial escape velocity of $v_{\rm esc} \sim 18\,\mathrm{km\,s^{-1}}$. We find that none of the 2G BHs are retained when the \HLZ{} model is used. In contrast, \gwModel{} yields a small population of retained 2G BHs. As before, no 2G+$n$G merger remnants are retained.

These results provide useful demonstrations of the impact of recoil kicks on hierarchical BBH mergers in clusters and motivates a comprehensive exploration across a wide range of cluster initial conditions, initial BH populations, and binary interaction prescriptions. Such an extended study is beyond the scope of the present analysis and will be pursued in a follow-up investigation.

\section{Discussion and Conclusions}
\label{sec:discussion}
In this paper, we performed a detailed study of hierarchical BH mergers in clusters using a newly developed and more accurate recoil-kick model for BBH mergers, \gwModel{}~\cite{Islam:2025drw}, and compared its results with the widely-used analytic model \HLZ{} \cite{Lousto:2008dn,Lousto:2010xk,Lousto:2012gt,Lousto:2012su,Gonzalez:2007hi}.

We used a variety of theoretical astrophysical toy models of star clusters to understand the recoil-kick model systematics and to address several astrophysically interesting questions, such as: How massive can BHs become purely through hierarchical mergers? How do the BH mass and spin distributions evolve across successive generations? How frequently are BH merger remnants retained within a cluster? How do the initial mass and spin distributions of 1G BHs and their binary pairing functions affect hierarchical-merger outcomes? These simplified toy models allow us to isolate and understand the impact of recoil-kick modeling choices without additional astrophysical complications. 
We then used a detailed semi-analytical framework for cluster evolution (\rapster{}) and also used previous simulation catalogs to investigate how more complex physical processes, such as cluster expansion, evaporation, binary hardening, and dynamical exchanges, modify hierarchical-merger pathways and their observable consequences.

We showed that, in most cases, the updated \gwModel{} prescription leads to the formation of more massive BHs through hierarchical mergers compared to the \HLZ{} model that has been employed in essentially all $N$-body and population-synthesis frameworks to date. Furthermore, \gwModel{} produces a significantly larger number of BHs that occupy the binary-parameter space corresponding to the progenitor BHs inferred for GW231123, whereas \HLZ{} struggles to populate this region in most cases.

As a by-product of our analysis, we also demonstrated that the initial BH population, the binary pairing function, and the prescription for binary hardening all have significant impacts on hierarchical-merger outcomes. This underscores the importance of performing multiple simulations that systematically vary these physical ingredients in order to fully capture the range of possible hierarchical-merger pathways.

Since our study highlights the importance of using accurate recoil-kick prescriptions in cluster-evolution modeling, it is natural to incorporate such prescriptions into other semi-analytical cluster-evolution frameworks, such as \texttt{fastcluster}~\cite{Mapelli:2021Gyv} and \texttt{BBHDynamics}~\cite{Antonini:2019ulv}, \texttt{BPOP}~\cite{Sedda:2021vjh}, as well as into direct $N$-body and Monte Carlo cluster codes, including \texttt{Cluster Monte Carlo}~\cite{Rodriguez:2019huv,Kremer:2019iul} \texttt{MOCCA}~\cite{2013MNRAS.431.2184G}, \texttt{SEVN}~\cite{Spera:2017fyx}, and \texttt{nbody6}~\cite{2015MNRAS.450.4070W}, \texttt{PeTar}~\cite{2020MNRAS.497..536W}. This is an avenue we will pursue in future work. We will also explore distilling our normalizing flow model for $v_\mathrm{kick}$ 
into a closed-form conditional distribution via symbolic regression. A compact 
analytic expression would be straightforward to incorporate into existing cluster 
simulation codes without external ML package dependencies (e.g., pytorch, nflows), while also reducing the per-sample computational cost relative to neural network inference.

Another important avenue for future work is to perform cluster-evolution simulations across a broader range of cluster properties using \gwModel{} in order to determine how hierarchical-merger pathways change relative to those predicted by \HLZ{}. It will be particularly interesting to assess whether this leads to a significantly larger population of massive BHs in clusters and thereby provides a more natural explanation for events such as GW231123. We will leave this investigation to future work.

\begin{acknowledgments}
We thank Emanuele Berti, Katie Breivik, Maya Fishbach, Norbert Langer, Carl Rodriguez, and Tejaswi Venumadhav for discussion. 
T.I. is supported in part by the National Science Foundation under Grant No. NSF PHY-2309135 and the Gordon and Betty Moore Foundation Grant No. GBMF7392. 
Use was made of computational facilities purchased with funds from the National Science Foundation (CNS-1725797) and administered by the Center for Scientific Computing (CSC). The CSC is supported by the California NanoSystems Institute and the Materials Research Science and Engineering Center (MRSEC; NSF DMR 2308708) at UC Santa Barbara. 
D.W. and K.K. are supported by NSF Grants No.~AST-2307146, No.~PHY-2513337, No.~PHY-090003, and No.~PHY-20043, by NASA Grant No.~21-ATP21-0010, by John Templeton Foundation Grant No.~62840, by the Simons Foundation [MPS-SIP-00001698, E.B.], by the Simons Foundation International [SFI-MPS-BH-00012593-02], and by Italian Ministry of Foreign Affairs and International Cooperation Grant No.~PGR01167. In addition, K.K. is supported by the Onassis Foundation - Scholarship ID: F ZT 041-
1/2023-2024. This work was carried out at the Advanced Research Computing at Hopkins (ARCH) core facility (\url{https://www.arch.jhu.edu/}), which is supported by the NSF Grant No. OAC-1920103. 
\end{acknowledgments}

\bibliography{references}

\appendix
\section{Extra Plots}
\label{sec:appendix}
\begin{figure}
    \centering
    \includegraphics[width=\columnwidth]{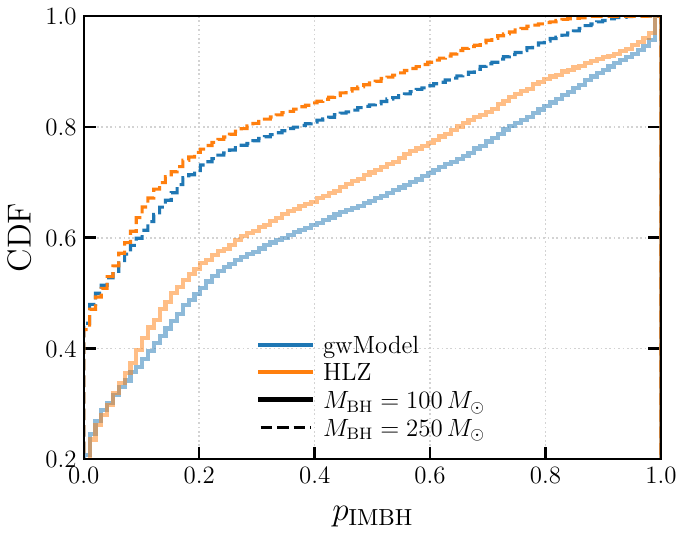}
    \caption{We show the cumulative probability of forming a black hole with a mass of $100\,M_\odot$ (a typical lower bound for an IMBH; solid lines) or $250\,M_\odot$ (roughly, the remnant mass of GW231123; dashed lines) through successive mergers for both the \gwModel{} (blue) and \HLZ{} (orange) prescriptions. The probabilities are computed for escape velocities of $200$ and $500~\mathrm{km\,s^{-1}}$ and for three metallicities, $Z \in \{0.0002,\,0.002,\,0.02\}$. Details are in Section~\ref{sec:pimbh}.}
    \label{fig:pimbh_histograms}
\end{figure}

We repeat the exercise presented in Section~\ref{sec:pimbh} for three other metallicity values, $Z \in \{0.0002,\,0.002,\,0.02\}$, and two escape velocities, $[200,500]~\mathrm{km\,s^{-1}}$. For each metallicity, we repeat the experiment described above using both the \HLZ{} and \gwModel{} prescriptions.
We plot the resulting overall probabilities (averaged over metalicities) of forming an IMBH with masses of $100\,M_\odot$ and $250\,M_\odot$ in Figure~\ref{fig:pimbh_histograms}.
We find that the \gwModel{} prescription consistently predicts a higher likelihood of IMBH formation than the \HLZ{} model. Overall, when \gwModel{} is used, the probability of forming an IMBH through successive mergers increases by approximately $5$--$8\%$.

\end{document}